\documentclass{aa}
\pdfoutput=1

\usepackage{graphicx}
\usepackage[varg]{txfonts}
\usepackage[]{hyperref}
\usepackage{multirow}  
\usepackage{enumitem}  
\usepackage{subfigure}
\usepackage{epstopdf}  

\usepackage{color} 

\newcommand{\teff}{\ensuremath{{T_{\rm eff}}}}           
\newcommand{\logg}{\ensuremath{\log g}}                  
\newcommand{\loggf}{\ensuremath{\log gf}}                
\def\kms{$\mathrm {km~s}^{-1}$}

\def\vsini {$v\sin i$}

\def\chisq{$\mathrm{\chi^2_{red}}$}
\def\liseven {$\mathrm{^{7}Li}$}
\def\lisix {$\mathrm{^{6}Li}$}
\def\iso{$\mathrm{^{6}Li/^{7}Li}$}
\def\1213C{$\mathrm{^{12}C/^{13}C}$}

\newcommand{\linfor}{{\sf Linfor3D}}
\newcommand{\nlte}{{\sf NLTE3D}}
\newcommand{\COBOLD}{{\sf CO$^5$BOLD}}
\newcommand{\cobold}{\COBOLD}
\newcommand{\lhdm}{{\sf LHD}}

\begin{document}

   \title{Lithium abundance and $^{6}$Li/$^{7}$Li ratio in the active giant HD\,123351}

   \subtitle{I. A comparative analysis of 3D and 1D NLTE line-profile fits}

   \author{
                A. Mott\inst{1},
                M. Steffen\inst{1},
                E. Caffau\inst{2},
                F. Spada\inst{1}
                \and K. G. Strassmeier\inst{1}
        }

   \institute{
                Leibniz-Institut f{\"u}r Astrophysik Potsdam, An der Sternwarte 16, 14482 Potsdam, Germany\\
                \email{amott@aip.de}
                \and
                GEPI, Observatoire de Paris, PSL Research University, CNRS, Place Jules Janssen, 92195 Meudon, France
        }

\authorrunning{A.~Mott et al.}

\titlerunning{Lithium in the active giant HD\,123351}

\date{Received DD MM YY/\,Accepted DD MM YY}

\abstract
{Current three-dimensional (3D) hydrodynamical model atmospheres together with detailed spectrum synthesis, 
accounting for departures from Local Thermodynamic Equilibrium (LTE), permit to 
derive reliable atomic and isotopic chemical abundances
from high-resolution stellar spectra.
Not much is known about the presence of the fragile \lisix\ isotope in evolved
solar-metallicity red giant branch (RGB) stars, not to mention its production in magnetically 
active targets like HD\,123351.}
{A detailed spectroscopic investigation of the lithium resonance doublet in 
HD\,123351 in terms of both abundance and isotopic ratio is presented. 
From fits of the observed spectrum, taken at the Canada-France-Hawaii telescope, with synthetic line profiles based on
1D and 3D model atmospheres, we seek to estimate the abundance of the \lisix\ 
isotope and to place constraints on its origin.}
{We derive the lithium abundance $A$(Li) and the \iso\ isotopic ratio by
fitting different synthetic spectra to the Li-line region of a
high-resolution CFHT spectrum ($R$=120\,000, S/R=400). The synthetic spectra
are computed with four different line lists, using in parallel 3D
hydrodynamical \cobold\ and 1D \lhdm\ model atmospheres 
and treating the line formation of the lithium components in non-LTE (NLTE).
The fitting procedure is repeated with different assumptions and wavelength
ranges to obtain a reasonable estimate of the involved uncertainties.}
{We find $A$(Li)$\mathrm{=1.69\pm0.11~dex}$ and $\mathrm{^6Li/^7Li=}$\,
 8.0\,$\pm$\,4.4\% in 3D-NLTE, using the line list of
\mbox{\citet{melendez12}}, 
updated with new atomic data for \ion{V}{i},  
which results in the best fit of the
lithium line profile of HD\,123351. Two other line lists lead to similar
results but with inferior fit qualities.}
{Our $2\,\sigma$ detection of the \lisix\ isotope is the result of a
careful statistical analysis and the visual inspection of each achieved
fit. Since the presence of a significant amount of \lisix\ in the atmosphere
of a cool evolved star is not expected in the framework of standard stellar
evolution theory, non-standard, external lithium production mechanisms,
possibly related to stellar activity or a recent accretion of rocky
material, need to be invoked to explain the detection of \lisix\ in
HD\,123351.}

\maketitle


\section{Introduction}

All stellar evolution models predict a significant depletion of the surface
abundance of the fragile lithium isotopes $^6$Li and $^7$Li with time, because
mixing processes expose the bulk of the convective envelope, including the
surface layers, to the higher temperature of deeper layers: $^7$Li is
efficiently destroyed at temperatures $T \gtrsim 2.5 \cdot 10^6$ K, and $^6$Li
at even lower temperatures \citep{pinsonneault97}.

In standard stellar evolution models, the only source of mixing is convection.
In stars of mass 0.35\,$\lesssim M/M_\odot$\,$\lesssim$ 1.3, 
the bottom of the
outer convection zone begins to recede from the center towards the surface
during the late pre-main sequence (pre-MS).  As a consequence, lithium depletion
eventually stops as the temperature at the base of the convection zone becomes
too low for significant nuclear burning of this element.  Standard models
therefore predict significant lithium depletion during the pre-MS
phase, whereas little or no depletion is expected during the main sequence
\citep{iben65,forestini94}.  For this reason, the standard models 
overestimate the present-day solar lithium abundance by about a factor 
of $\approx 100$, and are at odds with the main sequence lithium depletion 
pattern observed in open clusters (e.g., the Pleiades, the Hyades, and M67, 
see \citealt{somers14} and references therein). This discrepancy is usually 
reconciled by introducing into the models non-standard sources of extra mixing, 
for example, rotationally induced, and/or related to early mass loss \
\citep{schatzman77,chaboyer95,eggenberger10}.

During the post-main sequence, Li depletion resumes as the star approaches the
red giant branch (RGB). As the convection zone deepens, both further
destruction and dilution of the residual Li occurs due to convective
mixing. A star located at the ascent of the RGB possesses already a
significantly expanded convective envelope.
Consequently, not much surface lithium should be left on a giant star, and
therefore practically no \lisix, being the more fragile of the two isotopes.
However, \citet{sackmann99} demonstrated that fresh \liseven\ can be created
by means of the Cameron-Fowler mechanism \citep{cameron71} not only in massive asymptotic giant branch
(AGB) stars but also in low-mass red giants on the RGB. This requires deep extra 
mixing by circulation below the base of the standard convective envelope,
which transports the products of p-p and CNO nuclear processing from different 
parts of the hydrogen burning shell to the stellar surface layers 
\mbox{\citep[see also][]{Denissenkov2004}}.
This so-called cool-bottom processing can also reproduce the trend
with stellar mass of the \1213C\ observations in low-mass red giants
(\citealt{charbonnel98}, \citealt{boothroyd99}).

Not much is known about the \iso\ (and the \1213C) isotopic ratio in evolved 
stars, but knowledge of this plays an important role in stellar evolution 
\citep{lambert81} and helps to constrain or even exclude some mixing mechanisms
\citep{charbonnel00}. 
In magnetically active stars, extra \liseven\ (and \lisix) is likely produced
in energetic flares due to accelerated $^3$He reactions with $^4$He
\citep{montes98,ramaty00}.  Investigating a possible connection between the 
lithium abundance of a star and its level of magnetic activity is evidently 
worthwhile.

The measurement of isotopic ratios from stellar spectra is more challenging than
the normal lithium and carbon abundance determinations because, for example,
the less abundant \lisix\ isotope in the stellar photosphere only manifests
itself as a subtle extra depression in the red wing of the lithium resonance 
doublet. Very high quality, high-resolution data and reliable model 
atmospheres are mandatory ingredients for such an analysis of the lithium 
isotopic ratio. Although the use of three-dimensional (3D) model atmospheres 
and line formation in non-local thermodynamic equilibrium (NLTE) is 
computationally demanding, it has recently become a viable approach in 
studying isotopic abundances in cool stars \citep[e.g.,][]{steffen12,lind13}
and is usually preferred to the local thermodynamic equilibrium (LTE) assumption.

\object{HD\,123351} is a K0 RGB star on its first ascent on the RGB, 
showing a strong \ion{Li}{i} feature at \mbox{670.8 nm}, indicating a
significantly higher lithium abundance than in the Sun.  The star is
magnetically active and is a component of a binary system in a highly
eccentric orbit ($e=0.81$) with an orbital period of 148\,d
\citep{strassmeier11}. Its rotation period is 58\,d and thus not synchronized
to the orbital motion (rotating five times faster than the expected
pseudo-synchronous rotation rate). This situation leaves room for the
speculation that some non-standard lithium production and/or dredge-up
mechanism, whose role is still not fully understood, may be at work.  A
possible production channel supporting the presence of both \lisix\ and
\liseven\ in evolved post-main sequence objects is represented by energetic phenomena
occurring on the surface of magnetically active stars, such as stellar flares
\mbox{\citep{tatischeff07}}. In addition, extra dredge-up by enhanced mixing 
during the periastron passages in eccentric close binaries might give rise to
enhanced (non-standard) lithium abundances by cool-bottom burning via the
Cameron-Fowler mechanism (see above). Alternatively, the detection of
\lisix\ might indicate the existence of an extrasolar planetary system, part
of which has been accreted during the expansion of the red giant's envelope
and thus contaminated its atmosphere with \mbox{\lisix-rich} material (e.g.,
\citealt{siess99}, \mbox{\citealt{israelian01,israelian03}}).

In the following, Section~\ref{S2} describes our target star,
Section~\ref{S3} presents our inventory of the different blending line 
lists and their impact on the measurable Li feature, Section~\ref{S4} 
describes the computation of model atmospheres and the spectrum synthesis 
applied in this paper, whose results are presented in Sect.~\ref{S5}. 
We discuss possible interpretations of the \lisix\ detection in 
Sect.~\ref{S6}, while our conclusions are summarized in Sect.~\ref{S7}.

\section{HD\,123351 -- HIP\,68904}\label{S2}
\subsection{Properties of the target star}

\begin{figure}[t!]
        \centering
        \resizebox{\hsize}{!}{\includegraphics[clip=true]{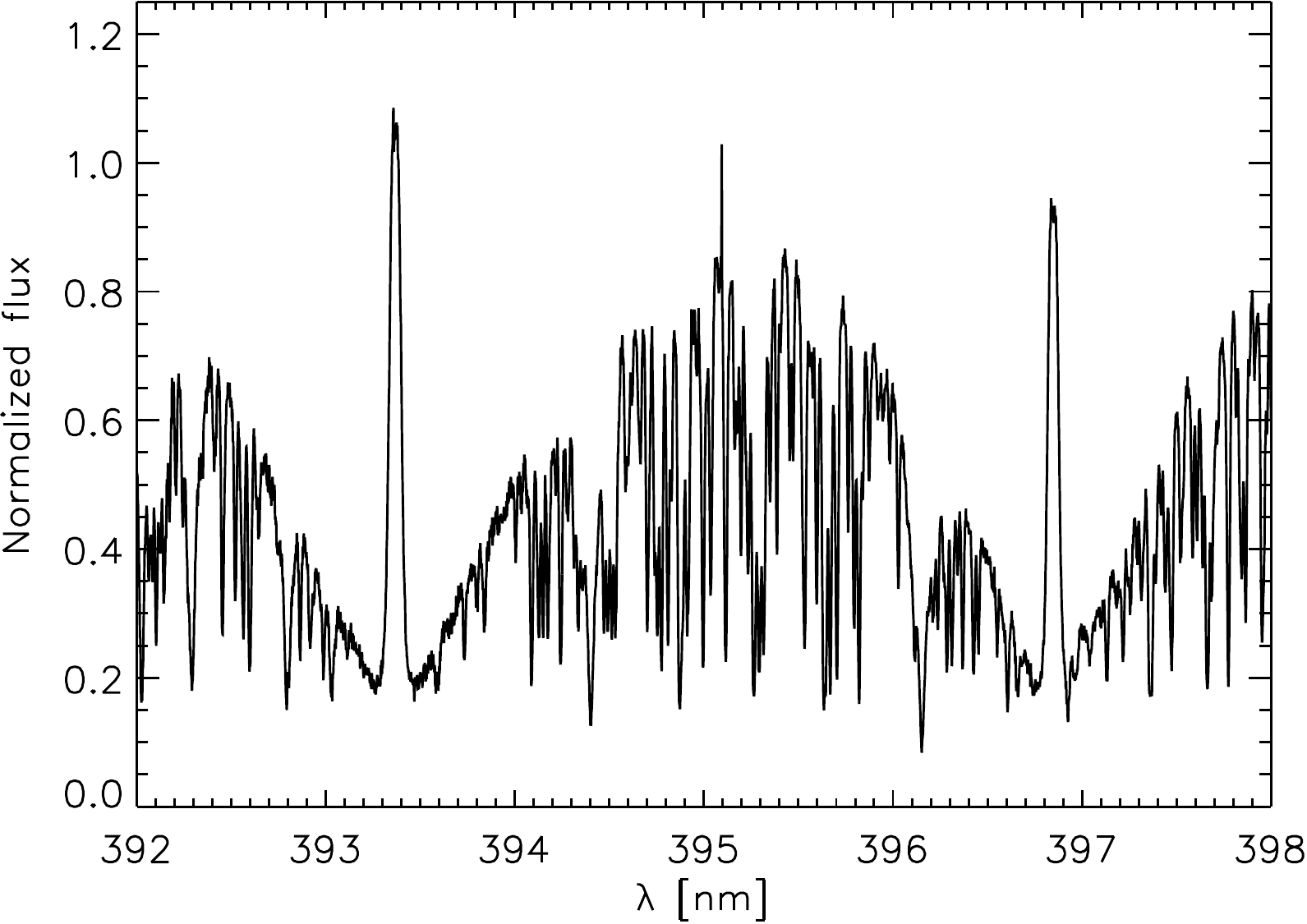}}
\caption{\ion{Ca}{II} H\&K lines in HD\,123351, taken from a SOPHIE spectrum 
with $R$=40\,000. The emission lines in the cores of the strong resonance lines 
are evident, revealing this star to be magnetically very active.}
\label{CaHK}
\end{figure}

The target is a recently studied example of a moderately Li-rich K0\,III-IV
giant \citep{strassmeier11}. It is heavily covered by spots and exhibits a
disk-averaged surface magnetic field of $\approx$540\,G.
This star was previously found to exhibit strong \ion{Ca}{ii} H\&K core
emission \citep{strassmeier00} which is known to be a fingerprint of
chromospheric magnetic activity. In \mbox{Fig.~\ref{CaHK}} we show a new
spectrum of the \ion{Ca}{ii} H\&K lines, part of a
SOPHIE\footnote{http://www.obs-hp.fr/guide/sophie/sophie-eng.shtml.}
high-resolution \'echelle spectrum (resolving power \mbox{R=40\,000},
signal-to-noise ratio \mbox{S/R=300} around $\lambda$670.8 nm), taken in
March~2014.

The HD\,123351 spectrum is characterized by a very low projected rotational
velocity (\mbox{\vsini\,=\,1.8 \kms}) and a prominent lithium line at
\mbox{670.8 nm} \mbox{($\mathrm{EW_{Li}\approx 73\,m}$\AA)}.  The one-dimensional (1D)-NLTE analysis of
\mbox{\citet{strassmeier11}} led to a lithium abundance of
\mbox{$A\mathrm{(Li) \footnotemark=1.70\pm0.05}$ dex}\footnotetext{$A\textrm{(Li)}=\log (N\textrm{(Li)}/N\textrm{(H)})+12$.},
but without information about a potential
contribution from the \lisix\ isotope. As mentioned above, standard stellar
models are not able to explain a significant \lisix\ content in RGB stars.
They find that this isotope is completely destroyed during the early stages of
evolution (pre-MS phase) as shown in \mbox{Fig. \ref{HR}b}.  On the other
hand, its presence was suggested by \mbox{\citet{strassmeier11}} who noted an
enhanced asymmetry of the Li line profile, which was found to be broader than
expected for the low \vsini\ of HD123351.  A detailed investigation of the
lithium resonance doublet through NLTE spectral synthesis is presented in this
work, with the parallel use of classical 1D and dedicated 3D hydrodynamical
model atmospheres.

\subsection{Comparison with stellar evolution models}
\label{S22}
\begin{figure}[t!]
{\bf a.}\\
\includegraphics[angle=0,width=0.5\textwidth, clip]{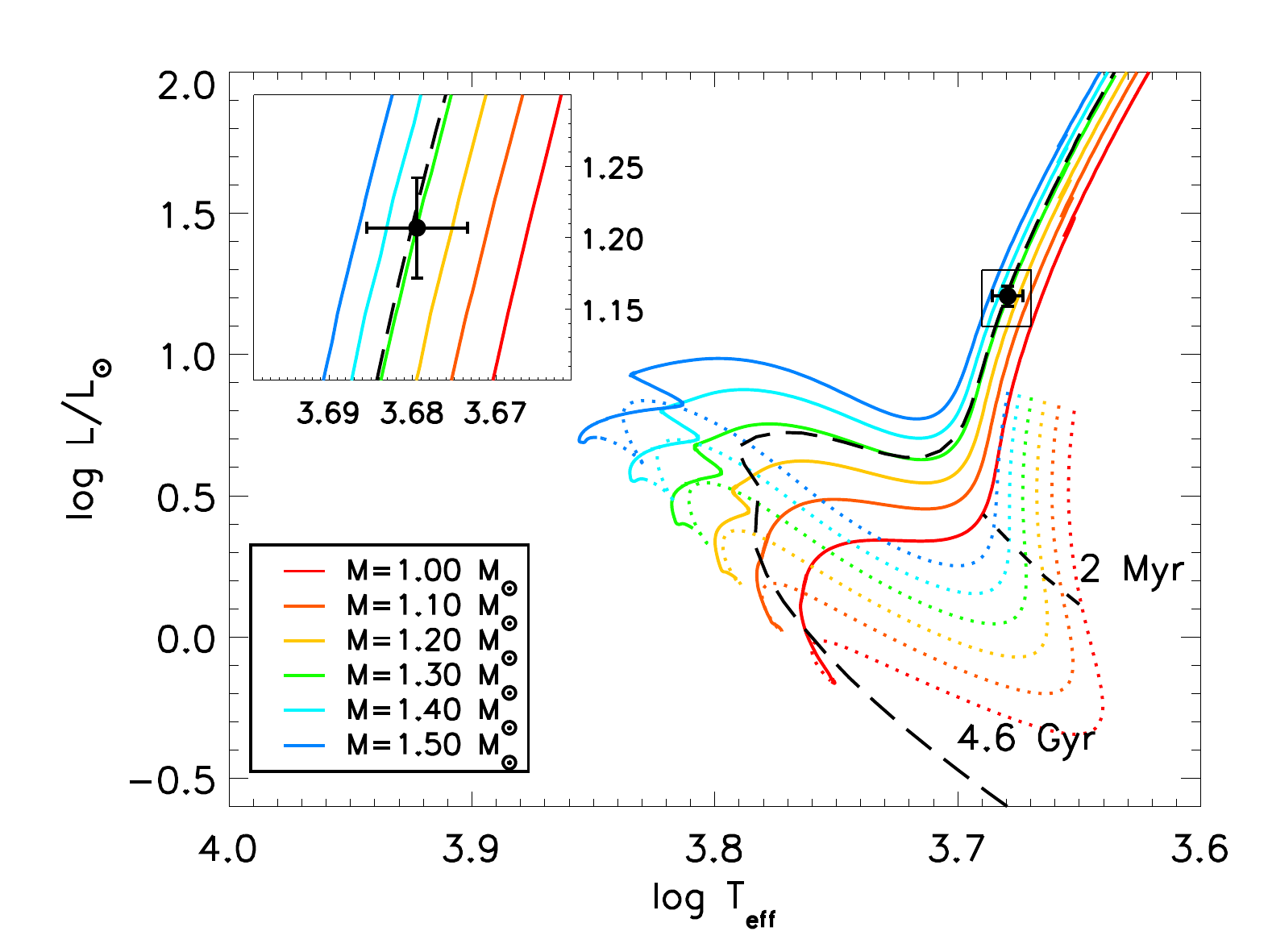}
{\bf b.}\\
\includegraphics[angle=0,width=0.5\textwidth, clip]{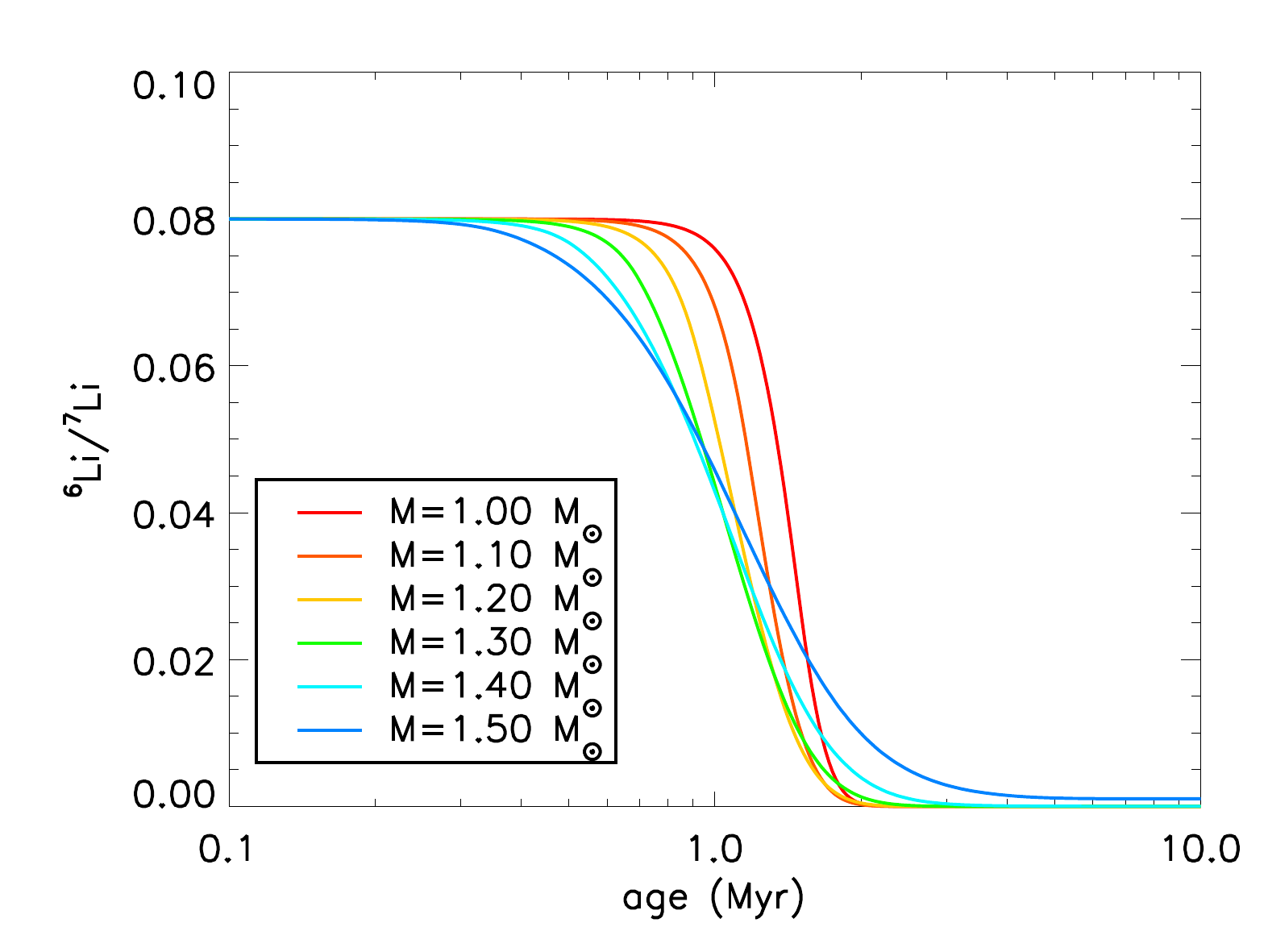}
\caption{\emph{a.} 
Position of the target star in the HR diagram compared with evolutionary
tracks and isochrones of solar metallicity. Tracks are shown in dotted
(pre-MS) and solid (main sequence onwards) colored lines;
isochrones at $2$ Myr and $4.6$ Gyr are plotted as short-dashed and
long-dashed black lines, respectively.
\emph{b.} \iso\ ratio as a function of age for the same models showing 
that \lisix\ is completely destroyed within approximately the first 
two million years of pre-MS evolution.}
\label{HR}
\end{figure}

The position of the target star in the HR diagram, together with evolutionary
tracks and isochrones of appropriate (solar) metallicity, are shown in
\mbox{Fig. \ref{HR}a}.  The tracks were constructed using the Yale Rotational
stellar Evolution Code (YREC) in its non-rotational configuration (see, e.g.,
\mbox{\citealp{demarque08}}), assuming a solar chemical composition
\mbox{\citep{grevesse98}} and a solar-calibrated value of the mixing length parameter
($\alpha_{\rm MLT}$\,=\,$1.82$). Microscopic diffusion of helium and heavy elements 
(not including Li) is
taken into account whereas convective core overshooting is not included in the
models, not to mention any other non-standard mixing processes.

We adopted the values of metallicity, effective temperature, and luminosity
from Table\,3 of \citet{strassmeier11}; the luminosity has been slightly
revised in consideration of the new parallax now available from the first Gaia
data release \citep{gaia}, which leads to smaller error bars on this quantity
($L/L_\odot = 16.1 \pm 1.3$).  The best-fitting YREC track has a mass of 
$1.3\,M_\sun$, from which we derive an age of $\approx 5$ Gyr, slightly younger
than the \mbox{6-7 Gyr} range given by \citet{strassmeier11}, relying on 
evolutionary models with core overshooting.

Our standard models predict that $^6$Li is completely destroyed 
early in the pre-MS, within the first two million years.  
As Figure \ref{HR}b shows, this result is quite robust with respect to varying 
the stellar mass within a reasonable range. 

Turning to \liseven, we obtain $A(^7\rm Li)\approx1.2$ at the age of $4.6$ Gyr 
for the $1.3 \, M_\odot$ model (for reference, our adopted initial isotopic 
lithium abundances are $A(^6\rm Li)=2.2$, $A(^7\rm Li)=3.3$; cf. 
\citealt{andersgrevesse89}). 
As mentioned before, the only source of mixing in our models is convection;
as a consequence, our predicted lithium abundance will be larger, in
  general, than that obtained with models including additional sources of
  mixing (for example, due to rotation).  The value of \liseven\ predicted by
  our standard models should therefore be considered as an upper limit 
 (see Sect. \ref{S6} for further discussion of this issue).

\subsection{Observations}

The spectrum used for the present Li study was taken at the
Canada-France-Hawaii telescope (CFHT) with the coud\'e echelle spectrograph
(\textit{Gecko}) and is the same as analyzed in \citet{strassmeier11}. It
consists of three consecutive exposures taken on one night in May 2000. The
average-combined CFHT spectrum is characterized by a resolving power of
R=120\,000 and a peak signal-to-noise ratio (S/R) of 400:1 per resolution element. The very high
quality of this spectrum allowed us to perform a detailed spectral analysis of
the \ion{Li}{i} resonance feature in terms of both abundance and
\iso\ isotopic ratio.

\section{Atomic and molecular data}\label{S3}

The main difficulty in deriving atomic and isotopic lithium abundances in a
star with solar-like metal content is represented by the blends arising from
various atomic and molecular lines overlapping with the lithium feature. A
complete laboratory atlas of the lines that populate this spectral region
around \mbox{670.8 nm}  still does not exist. An example for possible
artifacts is the fictitious line at \mbox{670.8025 nm}, first noted in the
solar spectrum by \mbox{\citet{mueller75}}, and subsequently attributed to
different elements by different authors without reaching a definite
conclusion. It is thus of general interest to identify an optimal line list
that is able to fit satisfactorily the lithium doublet together with the
blends from the other chemical species.

We address this idea by analyzing HD\,123351 with four lists of blending
lines, some of which have been previously used in lithium-related work:
\citet{reddy02}, \citet{ghezzi09}, \citet{melendez12} and Israelian (2014,
priv. comm.) (hereafter R02, G09, M12 and I14, respectively). To each set of
blend lines we added the lithium atomic data as elaborated by
\mbox{\citet{kurucz95}} which we simplified to 12 components that account for
the full lithium isotopic and hyper-fine splitting. A table of the lines
of each line list, and the list of lithium components used in this work are
given in Appendix~\ref{appendix:A}. The damping parameters were taken from
the \mbox{VALD-v3} database \mbox{\citep{kupka11}}. 

\begin{figure*}[!htb]
        \centering
        \resizebox{\hsize}{!}{\includegraphics[clip=true]{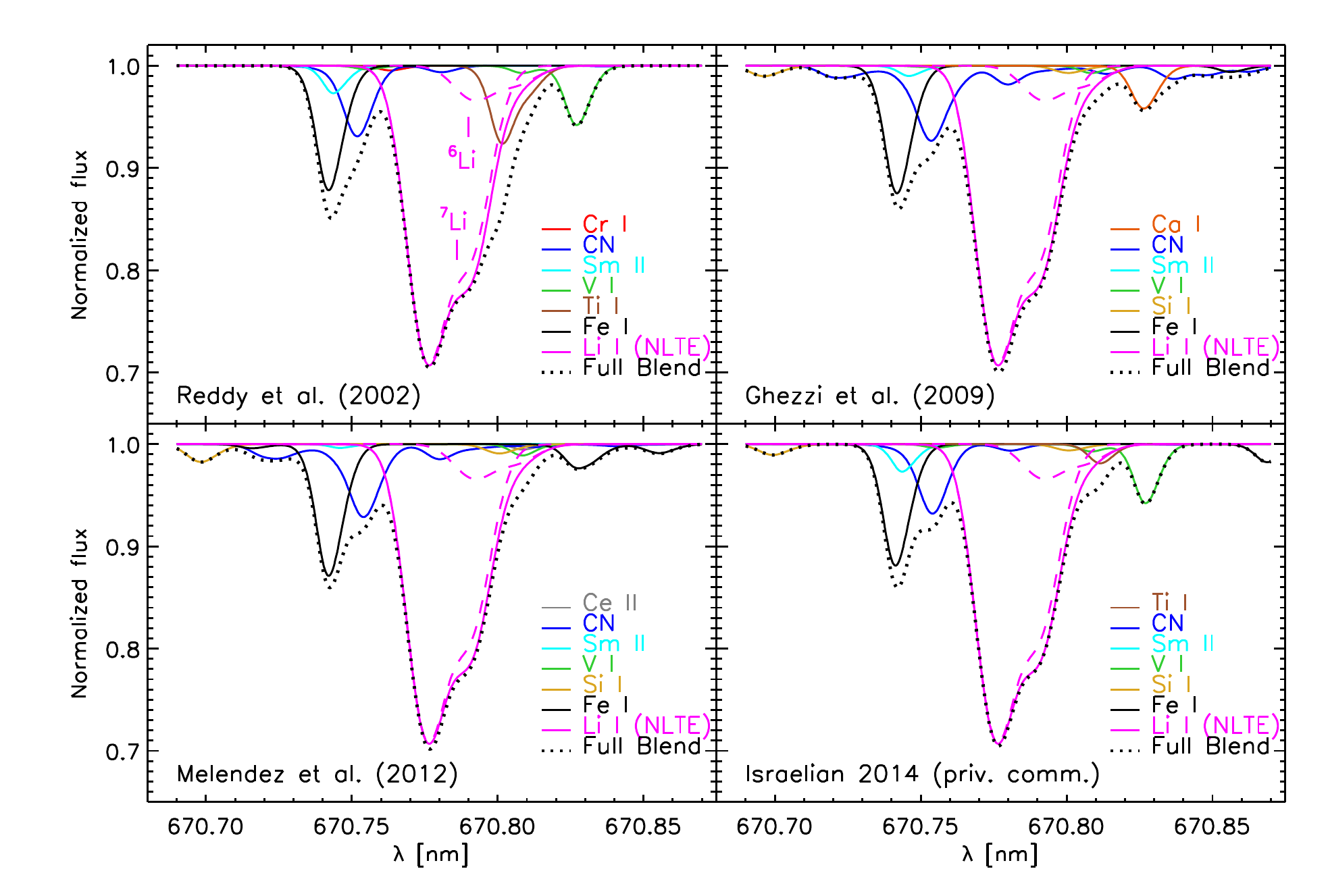}}
\caption{
Visualization of the four line lists used for the grid of synthetic Li line
profiles (flux). Each chemical species is synthesized in 3D separately and 
shown in a different color for easy identification.  We adopted model \texttt{A} of
\mbox{Table \ref{models}} with parameters closest to what was found by
\mbox{\citet{strassmeier11}}. The lithium line (magenta, dashed lines 
showing the contributions due to \lisix\ and \liseven) is computed in NLTE,
whereas we assumed LTE for the blend lines. The full blend is shown as the
dotted black line. No further broadening, except the 3D hydro-velocity field
arising from the convective motions, has been applied to the spectra. The line
lists differ in wavelength positions and \loggf-values, and in some cases even
in the chemical element assigned to a particular line, especially in the
critical red wing of the Li doublet where the \lisix\ components lie.
}
\label{linelists}
\end{figure*}

Figure~\ref{linelists} visualizes the main differences between the four
line lists. We plot the individual line contributions, computed separately 
for each element, so that their influence can be compared with the resulting full Li 
blend (dotted black line in \mbox{Fig. \ref{linelists}}).
For this exercise, we adopted model \texttt{A} from \mbox{Table \ref{models}} 
for all four panels, that is, the 3D-NLTE synthetic Li-line profile 
(magenta in \mbox{Fig. \ref{linelists}} for $A$(Li)\,=\,1.65 and 
\iso\,=\,0.082), and 3D-LTE profiles of the blend lines (color coded for 
each element). We note the different line identifications in the
red wing of the Li doublet where the blends interfere with the 
\lisix\ components.

\section{Stellar atmospheres and spectrum synthesis}\label{S4}

\subsection{3D \cobold\ and 1D \lhdm\ model atmospheres}

We utilized a sub-set from the CIFIST 3D hydrodynamical model atmosphere grid
\citep{ludwig09} computed with the \cobold\ code \mbox{\citep{freytag12}},
specifically tailored around the stellar parameters of HD\,123351. These
models are summarized in \mbox{Table \ref{models}}. The adopted values of \teff, \logg\ , and [Fe/H]
of model \texttt{A} are in accordance with what was derived by
\mbox{\citet{strassmeier11}}, whereas models \texttt{B} and \texttt{C} allow us to estimate the
sensitivity of the derived $A$(Li) and \iso\ ratio to \teff\ and \logg\ 
of the input model atmosphere. For each of these models, we also 
computed the corresponding 1D \lhdm\ model
\mbox{\citep{caffauludwig07}} (labeled as \texttt{a}, \texttt{b,} and
\texttt{c} respectively). In addition to the stellar parameters of the
\cobold\ models, the \lhdm\ models also share the micro-physics 
(opacity table, equation-of-state) and the numerics involved in treating
the radiative-transfer. As such, these 1D models are differentially 
comparable to the related 3D hydrodynamical models.

\begin{table*}[htbp]
\caption{List of 3D \cobold\ models and associated 1D~\lhdm\ model atmospheres 
used for the analysis of HD\,123351. }
        \label{models}
        \begin{tabular}{lcllcccccc}
        \hline\hline\noalign{\smallskip}
        Model name              &Label & Type & \teff    & \logg & $\mathrm{[Fe/H]}$  &      Box size       & Geom. size        & \#    &\#           \\
                                && & [K]   & cgs         & dex                &  X$\times$Y$\times$Z  & X$\times$Y$\times$Z [Mm]& Snaps &Bins \\
\noalign{\smallskip}\hline \noalign{\smallskip}
        \rule{0pt}{2.0ex}d3t48g32mm00n01 & \texttt{A} & 3D &4777$\pm$10&3.20&0.00&200$^2\times$140&109.7$^2\times$35.2 &20&5                    \\
        t4780g32mm00a05cifist & \texttt{a} & 1D LHD & 4780 & 3.20 &0.00 & -- & -- & -- & 5 \\
\noalign{\smallskip}\hline \noalign{\smallskip}
        \rule{0pt}{2.0ex}d3t46g32mm00n01  & \texttt{B} &3D & 4583$\pm$13&3.20&0.00&200$^2\times$140&109.7$^2\times$35.2 &22&5                                    \\
        t4580g32mm00a05 & \texttt{b} &1D LHD & 4580 & 3.20 & 0.00 & -- & -- & -- & 5 \\
\noalign{\smallskip}\hline \noalign{\smallskip}
        \rule{0pt}{2.0ex}d3gt46g35n03         & \texttt{C} &3D &4552$\pm$10&3.50&0.00&200$^2\times$140&109.7$^2\times$35.2 &12&5                                    \\
    t4550g35mm00ml3a05  & \texttt{c} & 1D LHD & 4550 & 3.50 & 0.00 & -- & -- & -- & 5 \\
    \noalign{\smallskip}\hline
\end{tabular}
\tablefoot{The last two columns denote the number of representative snapshots in the 3D models and the number of bins in which the opacities were grouped, 
following the opacity binning method (OBM, 
\citealt{nordlund82,ludwig94,vogler04}).}
\end{table*}

\begin{figure}[]
        \centering \resizebox{\hsize}{!}{\includegraphics[clip=true]{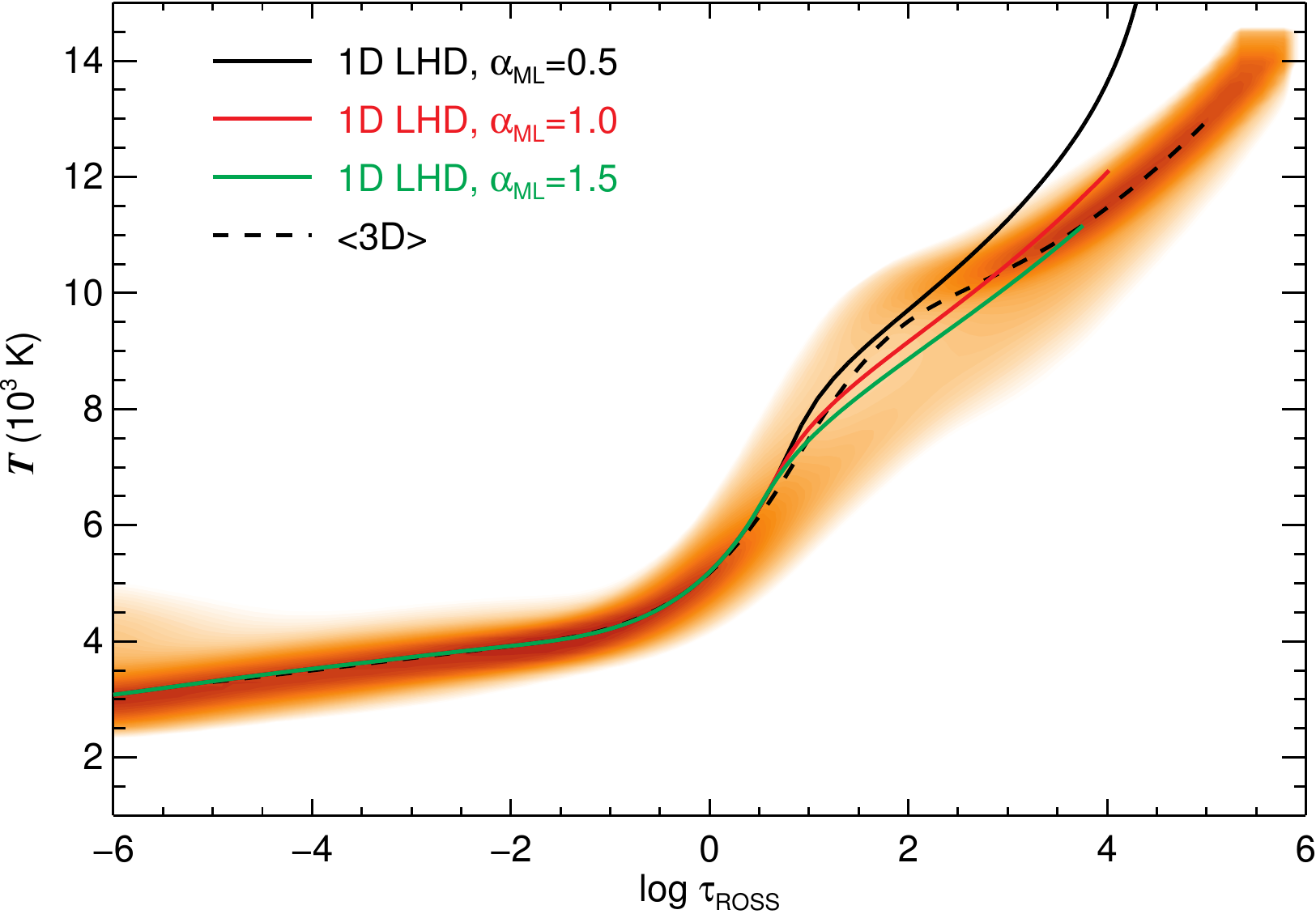}}
\caption{
Temperature structure of our 3D and 1D models on the Rosseland optical depth 
scale $\tau_{\mathrm{ROSS}}$. The adopted 3D model \texttt{A} (Table~\ref{models}) is shown
as a probability density distribution (orange). The 
$\langle\mathrm{3D}\rangle$ mean model is obtained by averaging the 20 
snapshots of the 3D model over surfaces of equal Rosseland optical depth 
(black dashed line). The three 1D \lhdm\ models with
$\alpha_{\mathrm{MLT}}=0.5$ (black), $1.0$ (red), and $1.5$  (green) are shown as
solid lines. The 3D model reveals photospheric temperature fluctuations in 
layers of fixed $\tau_{\mathrm{ROSS}}$ of several $100$\,K (darker shaded
regions are indicative of more likely temperatures).}
\label{tstruct}
\end{figure}

In Fig.~\ref{tstruct}, we show the thermal structure of the model atmospheres,
including the uppermost part of the underlying stellar convection zone. 
The temperature structure of the  adopted 3D model \texttt{A} is plotted as a
probability-density distribution in orange, while the 1D \lhdm\ models with
three different mixing-length parameters (\mbox{$\mathrm{\alpha_{MLT}}$}) are
shown as solid lines, and the averaged $\langle\mathrm{3D}\rangle$ model 
as a dashed line.
The latter is the result of a temporal and horizontal average of all the
snapshots of the 3D model over surfaces of equal Rosseland optical depth. 
By definition, this 1D model represents the average temperature structure 
$\langle T^4(\tau_{\rm ROSS})\rangle^{1/4}$ of the 3D model. Significant 
temperature deviations exist especially in the subphotospheric layers 
($\tau_{\rm ROSS} > 1$) where the role of \mbox{$\mathrm{\alpha_{MLT}}$} 
becomes relevant for the \lhdm\ models, leading to considerable temperature 
differences between the \lhdm\ model with \mbox{$\mathrm{\alpha_{MLT}=0.5}$} 
and the $\langle\mathrm{3D}\rangle$ atmosphere.

In the photosphere ($\tau_{\rm ROSS}\leq 1$), the temperature structures of 
the three \lhdm\ models are indistinguishable and almost perfectly coincide 
with the $\langle\mathrm{3D}\rangle$ model. As we shall
see in the next section, this is the line forming region of the Li resonance
line in HD\,123351, for which the choice of the mixing-length parameter is
thus completely irrelevant. In the 1D spectrum synthesis, we therefore fix 
the efficiency of convection arbitrarily to \mbox{$\mathrm{\alpha_{MLT}=0.5}$},
keeping in mind that adopting any other value would not affect the derived
$A$(Li) and \iso\ isotopic ratio.

\subsection{3D and 1D NLTE spectrum synthesis}\label{synthesis}

For each model and line list, a grid of synthetic spectra has been computed 
with the package \linfor\footnote{\url{http://www.aip.de/Members/msteffen/linfor3d}.} \citep{steffen15}
for a predefined sample of $A$(Li) and \iso\ values, covering the wavelength range
\mbox{[670.69\,--\,670.87] nm}.

\begin{figure}[]
	\centering
	\resizebox{\hsize}{!}{\includegraphics[clip=true]{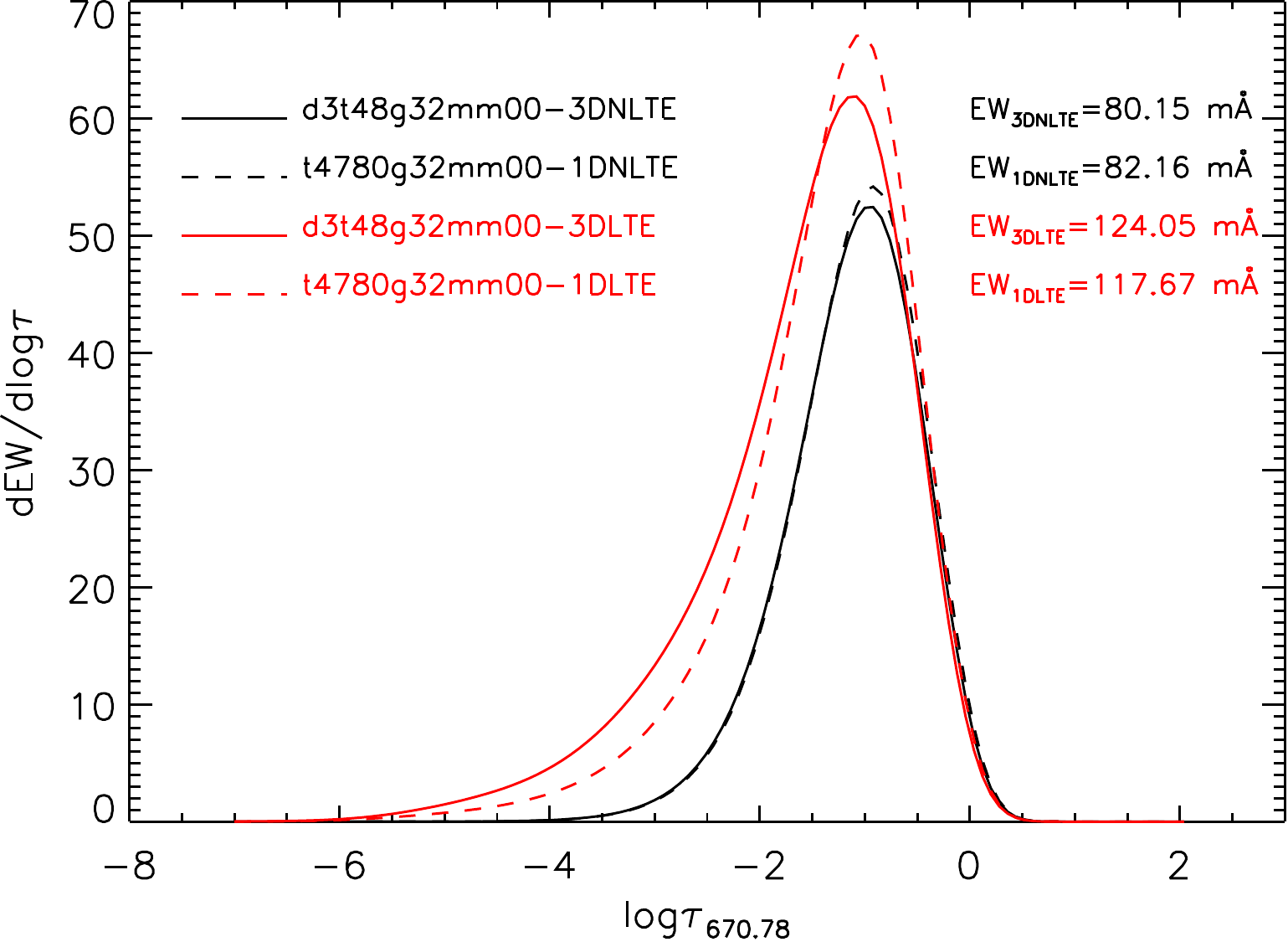}}
\caption{
Equivalent-width flux contribution function for the \ion{Li}{I} resonance
line. Shown are the results for the 1D~\lhdm\ model \texttt{a} (dashed curves, black 
for NLTE, red for LTE) and for the full 3D model \texttt{A} (interpolated to a common 
$\log\tau_{\mathrm{cont}}$ scale at 670.8\,nm, solid curves). The lithium
abundance is set to $A\rm(Li)=1.70$, and the parameter $\alpha_{\mathrm{MLT}}$ 
of the \lhdm\ model is 0.5.}
\label{cf}
\end{figure}

The micro-turbulence parameter required for the 1D line formation has been
derived from a set of \ion{Fe}{i} lines in the CFHT spectrum, such that any
correlation between reduced 1D-LTE equivalent widths and iron abundances
measured from these lines is eliminated. For HD\,123351 we found the value
\mbox{$\mathrm{\xi_{micro}=1.42}$ \kms} , which has been adopted in the 1D
spectrum synthesis. The rotational profile was applied afterwards by adopting
the fixed value of \mbox{\vsini=1.8 \kms} for both 3D and 1D fits
(\mbox{Sect. \ref{fitting_procedure}}).

The lithium line formation was treated in non-LTE, using a grid of departure 
coefficients precomputed for each Li abundance with the code \nlte. 
The code relies on an upgraded version of the Li model atom developed by 
\citet{cayrel07} and described in \citet{sbordone10}. Our updated model atom
consists of 17 energy levels and accounts for 34 bound-bound radiative
transitions. For a few test cases, we compared the results obtained with 
this model atom to those obtained with an even more recent lithium model 
atom, with 26 levels and 96 radiative transitions 
\citep[see details in][]{klevas16}, and did not find any appreciable 
differences in the line profiles. We therefore decided to proceed with 
the already initialized computation of the grid of synthetic spectra for 
HD\,123351 with the 17 level model atom.  

\citet{cayrel07} showed that NLTE effects are particularly
important for the lithium line formation in metal-poor stars where 
they strongly reduce the height range of line formation such that 
the Li 3D NLTE equivalent width (EW) is reduced by a factor of two compared
to the 3D LTE case. For the stellar parameters of HD\,123351, the
NLTE effects are smaller but still substantial. Over-ionization of the
ground level and, to a somewhat lesser extent, of the first excited level
of the \ion{Li}{i} resonance line is the dominant NLTE effect. The
reduced line opacity and increased line source function,
relative to the LTE case, both lead to a weakening of the line.
The magnitude of the (positive) NLTE abundance correction is very 
similar in 1D and 3D ($\sim$\,0.2\,dex) and is in good agreement with the 
results of \citet{Lind09}.

A useful way to check the impact of NLTE line formation is
to compare the so-called equivalent-width Contribution Function (CF) in LTE
and NLTE. This also allows us to clarify at which depth in the stellar atmosphere 
specific spectral lines are being formed; CF$(\tau)$ gives the relative 
contribution of each atmospheric layer to the equivalent width of the line.

For the 3D \cobold\ model \texttt{A} (Table \ref{models}) and the corresponding 
1D~\lhdm\ model, synthetic spectra were computed for the Li feature 
only, assuming both LTE and NLTE. In addition to synthetic line profiles, 
\mbox{\linfor} also provides the equivalent-width flux CF, 
following the formalism described in \mbox{\citet{magain86}}. 
Figure~\ref{cf} illustrates the result for the \ion{Li}{i} 
$\lambda$\,670.8\,nm line, assuming a logarithmic lithium
abundance of \mbox{$A$(Li)=1.70 dex} according to the 1D-NLTE 
value derived by \citet{strassmeier11}. The CF is
constructed in units of \mbox{d$\mathrm{EW}$/d$\log\tau$} over the continuum
optical depth ($\log\tau_{\rm cont}$) scale at 670.8\,nm, such that the
integration of the curve over $\log\tau_{\rm cont}$ gives the total
equivalent width of the line. In this way, the CF allows an immediate 
estimate of the strength of the lithium line in HD\,123351 together with 
an indication at which atmospheric depth the line is formed.

If we consider the range of optical depth including, say, 80\%\ of the
contribution function as the line forming region, we may conclude that the
lithium resonance line is mainly formed between
\mbox{$-3<\log\tau_{\mathrm{cont}}<0$} in NLTE. In the LTE case, there are
some layers above the stellar surface that contribute more to the formation of
the lithium line, causing the CF to move slightly higher in the
atmosphere. This causes an increase of the EW of the \ion{Li}{i} line of
roughly 55\%\ from \mbox{80.15\,m\AA} in NLTE, to 124.05\,m\AA\ in LTE
using the 3D hydrodynamical model. A similar change in EW is 
appreciated using 1D model atmospheres, for which neglecting departures 
from LTE leads to an EW that is 43\%\ larger with respect to the NLTE
assumption. 
In NLTE, the overall shapes of the CFs in 1D and 3D are quite
similar among each other, and we expect that 3D effects in HD\,123351 have 
only a small (but non-negligible) impact on the EW of the lithium line. 
However, the point of using 3D synthetic spectra is that the intrinsic 
shape of the 3D line profiles are supposedly more realistic (slightly 
asymmetric) than in 1D (intrinsically symmetric). We thus emphasize that 
the combined 3D NLTE effect, which can alter not only the strength but
also the shape and wavelength shift of a spectral profile by appreciable 
amounts, should be taken into account in the determination of reliable 
isotopic lithium abundances.

For synthesizing the blend lines in the lithium region, we have to assume
LTE, lacking the capabilities to take into account non-LTE effects for the
heavier elements with more complex atomic structures. We argue, however, 
that non-LTE effects of the blend lines are second-order corrections that
have only a minor impact on the analysis of the Li profile.

The abundance of all metals (excluding Li) is assumed to be solar
according to \citet{grevesse98}, except for C, N, and O, for which we
adopted the values recommended by \citet{asplund05}. A detailed
abundance analysis of the relevant chemical elements based on
recently observed SOPHIE spectra is presented in 
Appendix\,\ref{appendix:B1}. The results shown in Table\,\ref{tab:chem}
confirm that the chemical composition of HD\,123351 is consistent 
with the adopted solar abundance mix.

\subsection{Fitting procedure}\label{fitting_procedure}

We derived the lithium abundance $A$(Li) and the \iso\ isotopic ratio by 
fitting the CFHT spectrum of HD\,123351 with synthetic spectra obtained 
by interpolation from the pre-computed grid of synthetic Li line profiles.
For this purpose, we employ the least-squares fitting algorithm MPFIT
\mbox{\citep{markwardt09}} (implemented as an Interactive Data Language (IDL) routine described in more
detail in \mbox{\citealt{steffen15}}). It adjusts iteratively four fitting
parameters until the best fit (minimum $\chi^2$) is achieved. The set of free
parameters includes $A$(Li), \iso, a global wavelength adjustment 
($\Delta v$), and a global Gaussian line broadening ($FWHM$), 
which are applied in velocity space to the synthetic interpolated 
line profile to match the observational data as closely as possible.  
$FWHM$ represents the full width half maximum of
the applied Gaussian kernel and, in the 3D case, it includes the instrumental
broadening plus an additional broadening which might have been missing from
the fixed rotational broadening of \mbox{$v\sin i=1.8$ \kms}. In the 1D 
case, it also includes the fudge parameter known as macroturbulence
that is omitted in the 3D case where both micro and macro
velocity fields are already provided by the 3D model atmospheres. We relied on
the original normalization of the CFHT spectrum, fixing the level of the
continuum ($clevel$) to 1.00 throughout the fitting procedure.

We fit the Li doublet using two different wavelength windows:
\begin{itemize}[noitemsep,nolistsep]
\item \emph{Full range.} It covers the full synthesized spectral region between 670.69 and 
\mbox{670.87\,nm}, including all the lines belonging both to lithium and 
to the blends of each particular line list.

\item \emph{Li range.} It excludes from the fit the majority of the blends located outside the Li
line (which are instead only included in the \emph{Full range}). This 
restricted range ($\approx$\,670.76\,--\,670.82\,nm) is helpful in 
understanding to what extent the blend lines beyond the Li feature, 
in particular the dominant \ion{Fe}{i} line at 670.74\,nm, are responsible 
for any change in the resulting $A$(Li) and \iso\ ratio.
\end{itemize}

\section{Results}\label{S5}

\subsection{Evaluation of $^6Li/^7Li$ isotopic ratio}

We performed the $\chi^2$ analysis described in
\mbox{Sect. \ref{fitting_procedure}} to find the best fit between the
observed CFHT spectrum and the interpolated synthetic spectra for 
the whole set of model atmospheres (Table\,\ref{models}), line lists 
(Fig.\,\ref{linelists}), and different fitting setups 
(\emph{Full, Li range}). Focusing on the resulting $A$(Li) and
\iso\ ratio, we firstly found that, among the four line lists, R02 was 
not able to reproduce correctly the lithium line profile of HD\,123351 
using any of the fitting setups and model atmospheres, leading in general 
to overall bad-quality fits and even negative values for the \iso\ ratio. 
A negative content of \lisix\ would mean that the synthetic line profile 
is too depressed at the wavelength position of this isotope 
(\mbox{$\approx$\,670.81\,nm}), indicating a likely erroneous blend around 
the \lisix\ line in R02. Referring to \mbox{Fig.\,\ref{linelists}} this 
blend is obviously the relatively strong line of \ion{Ti}{i}, which is 
absent or dramatically weaker in the other line lists for which a better 
agreement with the data is found.

\begin{figure*}[]
        \centering
        \resizebox{\hsize}{!}{\includegraphics[clip=true,angle=180]{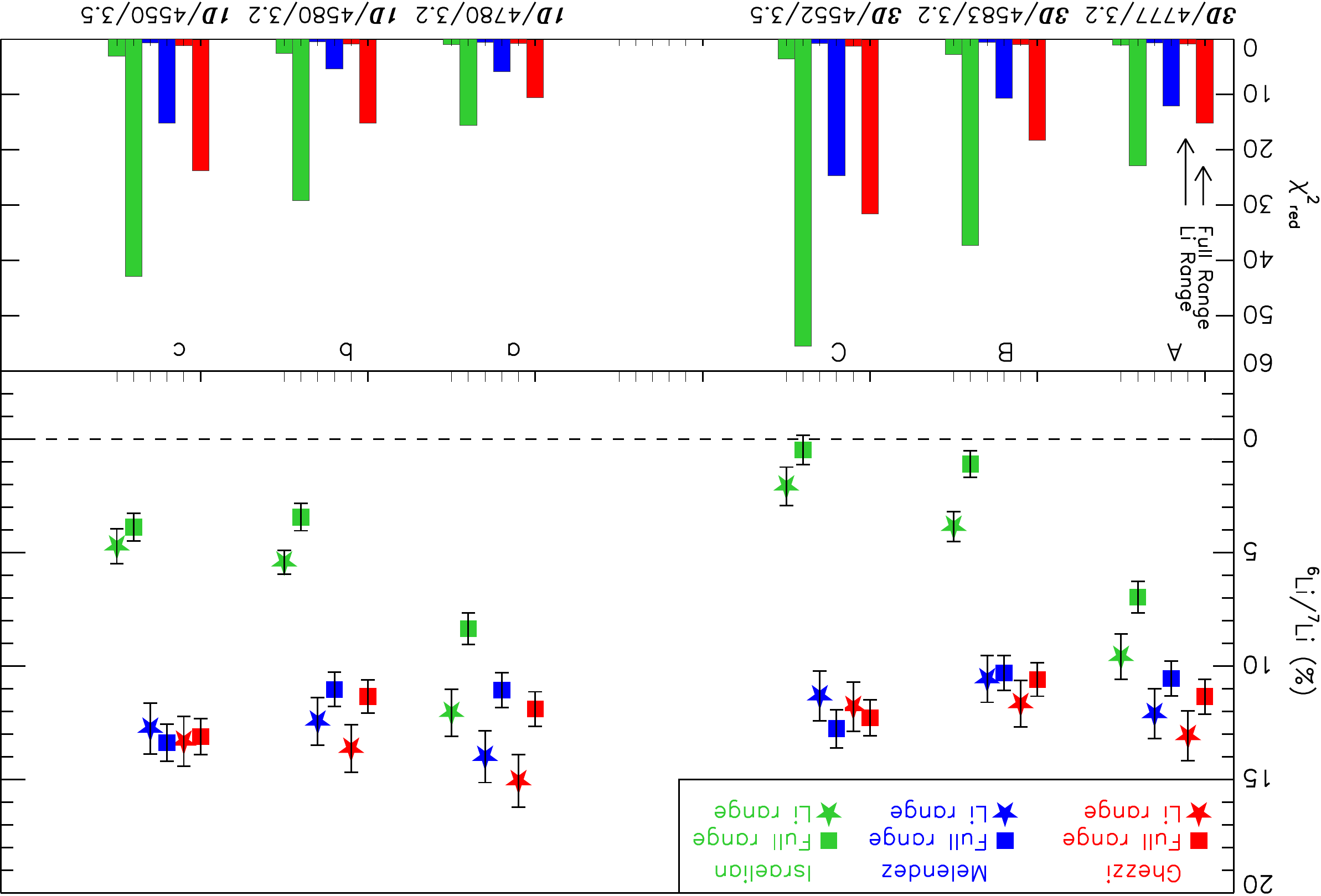}}
\caption{
Results of the analysis of the lithium resonance doublet in the CFHT spectrum 
of HD\,123351. The upper panel shows the \iso\ ratio for each line list derived
with 3D \cobold\ and 1D \lhdm\ model atmospheres (left and right part
respectively). The model parameters are indicated on the bottom axis.
Different symbols correspond to different fitting ranges; squares for the
\emph{Full range}, stars for the \emph{Li range}.  The error bars are the
formal 1$\sigma$ internal fitting errors on the free parameter \iso\ and do not
indicate the final uncertainty of the measurement. The quality of each fit is
shown in the lower panel where the bars represent the reduced $\chi^2$ value.}
\label{cfht_res}
\end{figure*}

We plot in \mbox{Fig. \ref{cfht_res}} 
the results for the isotopic ratios derived with all combinations of 
models and the three remaining line lists. The 3D model atmospheres are 
shown in the left half of the figure and the respective 1D~\lhdm\ models 
in the right half. The quality of each fit is expressed by the reduced 
\chisq\ in the lower panel.

Proving or disproving the detection of an isotope of very low abundance, such
as \lisix\ in a solar metallicity star, is a delicate task which requires a
pondered discussion. Referring to \mbox{Fig. \ref{cfht_res}}, we can
summarize the following findings.

\begin{itemize}[noitemsep,nolistsep]
\item
Using model \texttt{A}, we detect \lisix\ with all three line lists. Its value ranges
between 7\% and 13\% in 3D, depending on the adopted list of blend lines
and fitting range.

\item
With the 1D~\lhdm\ models, we find \iso\ to be systematically slightly larger
than in the 3D case. As suggested by \citet{cayrel07} 
\citep[see also][]{steffen10}, this is imputable to the fact that the 
asymmetry arising from convective motions is not modeled in 1D,
for which reason more \lisix\ is needed to reproduce the correct asymmetric
line profile\footnote{\citet{asplund06} found the opposite effect in the 
more complex case of using calibration lines to fix the rotational
broadening. We believe, however, that the use of independent calibration 
lines can easily lead to erroneous results, as argued in \citet{steffen12}.}.
Because of the more realistic treatment of the convection, the
results from 3D models are considered to be more reliable.

\item
The option of fitting the \emph{Li range} only (star symbols) provides, in
most cases, a \iso\ ratio up to 3\% larger than for the \emph{Full range}
(squares). This is an important effect caused by the \ion{Fe}{i} line at
\mbox{670.74 nm}. This line does not directly affect the lithium doublet in
HD\,123351 like, for example, a spurious blend; it rather intervenes, when included
in the fitting range, in shifting the synthetic profile towards the red to
better match the full blend, reducing the \lisix\ content needed to
reproduce the asymmetry. The \chisq\ obtained in fitting the \emph{Full
  range} is considerably larger due to the fact that the blends
outside the \emph{Li range} (especially the dominant \ion{Fe}{i} line) are not well
fitted.

\item
The line list of M12 seems to be the most suitable in fitting the full doublet
region of HD\,123351, providing \mbox{$A\rm (Li)=1.69~dex$} and \mbox{$\rm
  ^6Li/^7Li=10.56~\%$} using the model \texttt{A} with parameters closest to what was
found by \citet{strassmeier11}. Model \texttt{B} gives a fit of comparable quality
but finds a lithium abundance that is \mbox{0.25 dex} lower, as expected for
the lower \teff\ of this model.

\item
Using the line list of I14, we note a clear decline of \iso\ towards the models with lower
\teff\, both in 3D and in 1D. This is due to the fact that the temperature
sensitive \ion{Ti}{i} line, located close to the position of the
\lisix\ isotope (see \mbox{Fig. \ref{linelists}}), is stronger with the cooler
models \texttt{B} and \texttt{C}, leading to a smaller \lisix\ content.  This blend is not
present in M12 and G09, that both provide better fits (lower \chisq),
suggesting that the \ion{Ti}{i} blend in I14 is probably incorrect. However, 
should this line identification prove to be correct, it would make a reliable
detection of \lisix\ in cool solar-type stars extremely challenging.

\end{itemize}

\subsection{ Updated \loggf\ value for the \ion{V}{i} line at 670.81\,nm}
\label{sec:VI} 

Recently, \citet{lawler14} provided improved values of both oscillator 
strength and wavelength for, among the others, the \ion{V}{i} line that lies 
very close to one of the \lisix\ components and is present in all the line 
lists considered in this work (see Fig.\,\ref{linelists}).
These new parameters ($\lambda=670.81096$ nm, $\loggf = -2.63$) are
significantly different from the values in the lists of blends available 
in the literature and used in this work. Depending on the line list 
considered, the new oscillator strength is 0.3 to 0.5 dex larger 
(cf.\, Tables in Appendix\,\ref{appendix:A}).

To investigate how this affects our \lisix\ detection in HD\,123351, 
we updated the wavelength and \loggf\ value of the \ion{V}{i} line with 
the new measurements of \citet{lawler14} in all of our line lists.
After fully re-computing the grids of synthetic 1D and 3D NLTE spectra 
with the modified line lists, we proceeded with fitting the CFHT spectrum 
as described in Sect.\,\ref{fitting_procedure}. The results of the best fits 
obtained with model atmospheres \texttt{A} and \texttt{a} 
(Table \ref{models}) and line lists G09, M12, and I14, are
presented in Table \ref{VImod}, together with the results derived
with the original line lists for direct comparison.

\begin{table}[!htb]
\caption{Results of the NLTE best fits to the CFHT spectrum adopting the 
3D model \texttt{A} and the 1D \lhdm\ model \texttt{a}, the line lists of
\citet{ghezzi09}, \citet{melendez12}, Israelian (priv. comm. 2014), and 
the two versions of the \ion{V}{i} line at \mbox{$\sim$670.81 nm}.}
\label{VImod}
\centering
 \resizebox{\columnwidth}{!}{
\begin{tabular}{lcccccc}
\hline\hline\noalign{\smallskip}
\multicolumn{7}{c}{Original \ion{V}{i} line - \emph{Full range}}                        \\                         \hline\noalign{\smallskip}
& \multicolumn{3}{c}{3D model \texttt{A}}  & \multicolumn{3}{c}{1D model \texttt{a}}                  \\    \hline\noalign{\smallskip}
        & \textbf{G09}   & \textbf{M12}    & \textbf{I14} & \textbf{G09} &  \textbf{M12}& \textbf{I14}\\    \hline\noalign{\smallskip}
$A$(Li) &  1.688 &  1.689  &  1.701 & 1.671          &  1.672          & 1.683   \\ 
\iso    & 11.352 & 10.556  &  6.964 &11.898          & 11.065          & 8.354   \\ 
\chisq  & 15.178 & 12.082  & 22.828 &10.590          &  5.898          &15.589   \\                         \noalign{\smallskip}\hline\noalign{\smallskip}
\multicolumn{7}{c}{\ion{V}{i} line from \citet{lawler14} - \emph{Full range}}           \\                         \hline\noalign{\smallskip}
& \multicolumn{3}{c}{3D model \texttt{A}}  & \multicolumn{3}{c}{1D model \texttt{a}}    \\                  \hline\noalign{\smallskip}
        & \textbf{G09}   & \textbf{M12}    & \textbf{I14} & \textbf{G09} &  \textbf{M12} &\textbf{I14}\\    \hline\noalign{\smallskip}
$A$(Li) &  1.685 & 1.687   & 1.698 & 1.668          &  1.670          & 1.680    \\ 
\iso    &  7.816 & 7.984   & 3.624 & 8.889          &  8.863          & 5.523    \\
\chisq  & 15.578 & 11.856  &23.827 & 10.961         &  5.669          &16.358    \\                         \noalign{\smallskip}\hline
\end{tabular}
}
\end{table}

As expected, a stronger contribution from the \ion{V}{i} blend at this 
critical wavelength position has the natural effect of decreasing the amount 
of \lisix\ needed to best fit the data. For the line lists G09 and I14, the 
change in \loggf\ is roughly $+0.5$\,dex with respect to the original 
oscillator strength, and has the effect of lowering the lithium isotopic 
ratio by up to $\sim$3.5 percentage points. For M12, the smaller difference 
in the oscillator strength  ($\Delta\loggf$\,$=$\,$+0.3$\,dex) leads to a \iso\ 
ratio which is lower by 2.6 (2.2) percentage points in 3D (1D).
Due to the vicinity of this \ion{V}{i} line to one of the \lisix\ components, 
its impact on the \iso\ ratio is non-negligible.
On the other hand, the lithium abundances $A$(Li) obtained with the 
modified \ion{V}{i} line are basically unchanged for the three lists of 
blend lines. We also note that the modified line list M12 is still the 
best choice to reproduce the lithium doublet region around 670.8\,nm, 
as indicated by the \chisq\ values listed in Table\,\ref{VImod},
while the update of the \ion{V}{i}  line does not significantly improve
the quality of each best fit.

\begin{figure*}[!htbp]
{\bf a. \hspace{92mm} b.}\\
\mbox{\includegraphics[angle=0,width=90mm, clip]{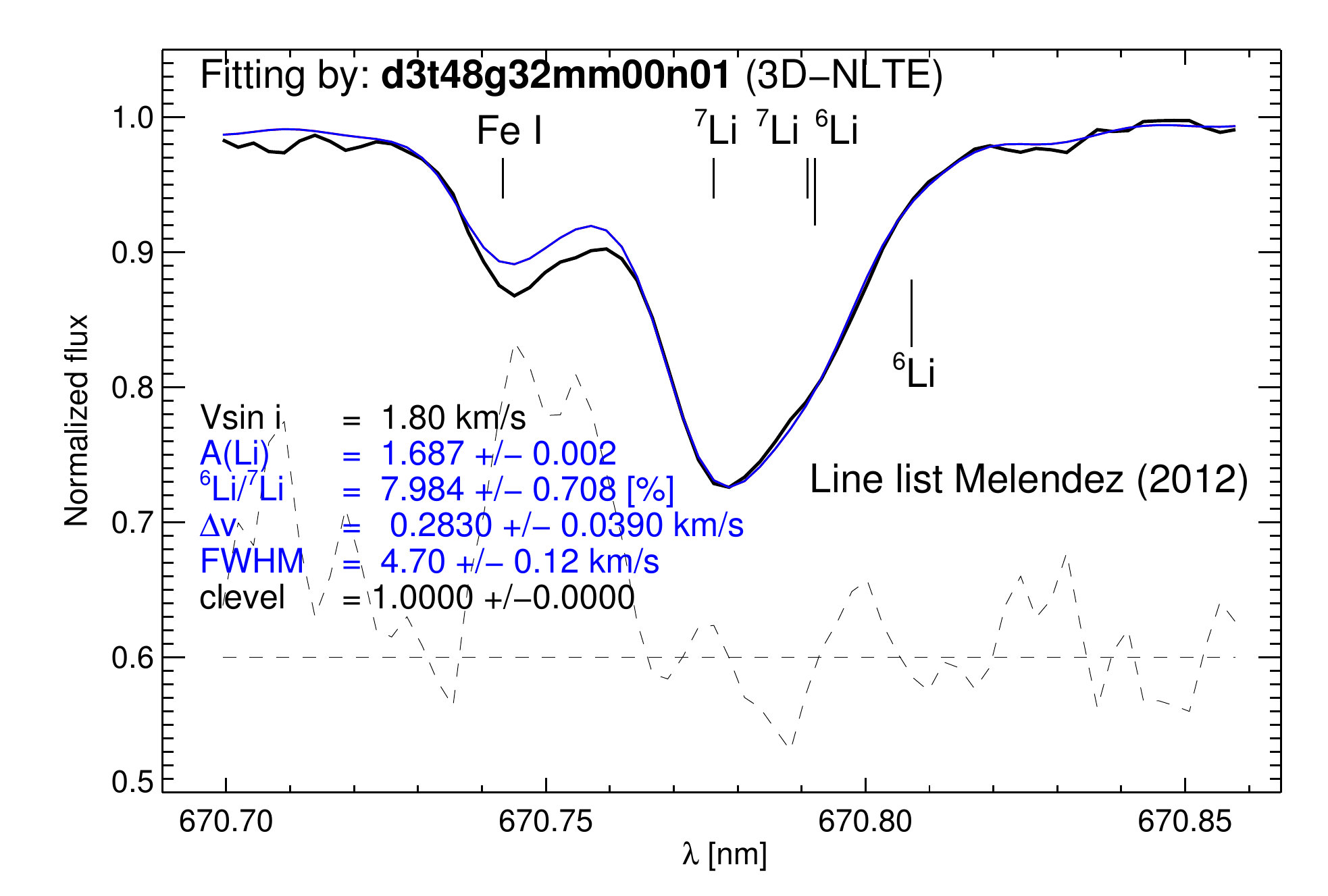}}
\hspace{0.00cm}
\mbox{\includegraphics[angle=0,width=90mm, clip]{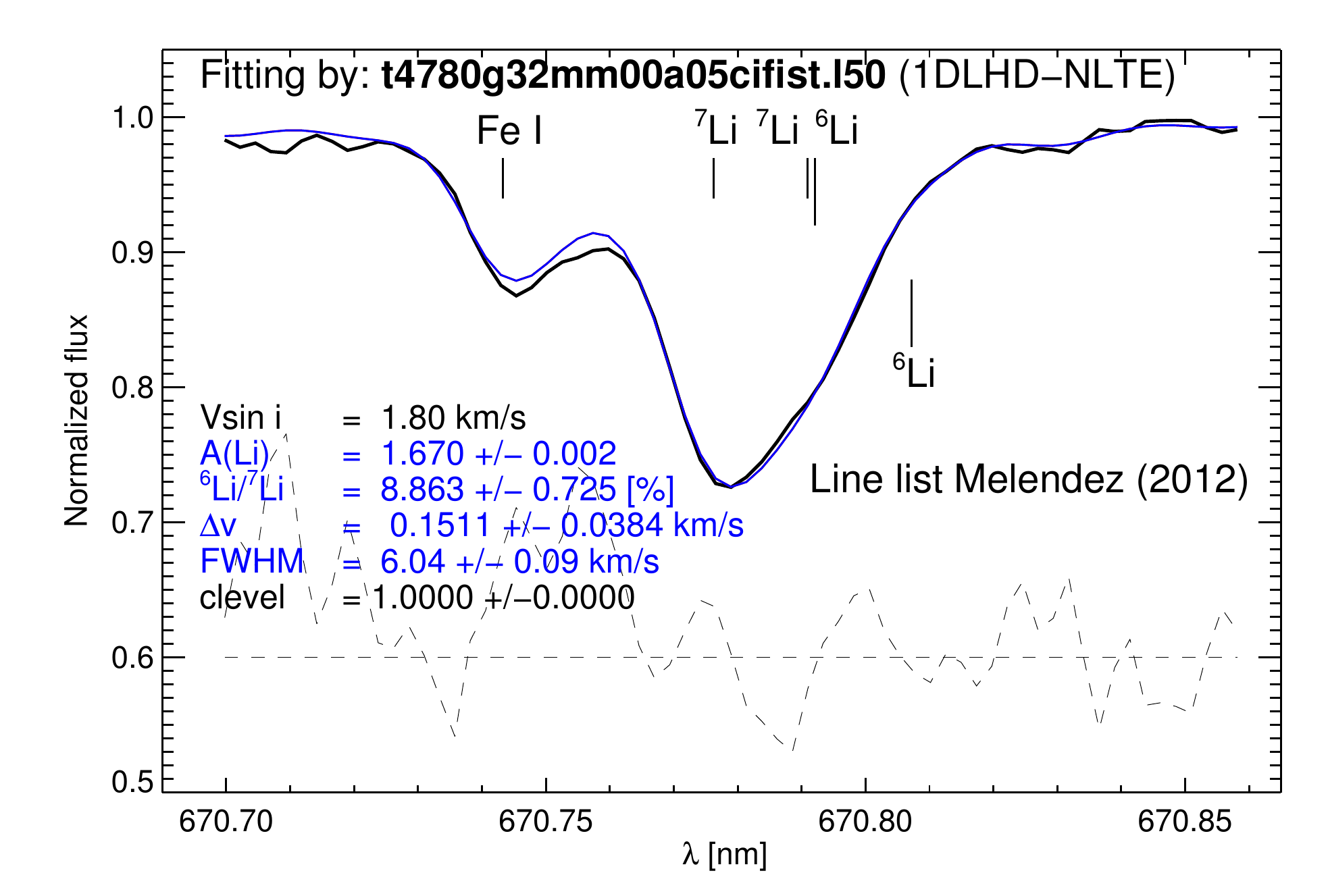}}
\caption{
Best fit (thin blue line) to the CFHT spectrum (black line) obtained with
line list M12+L14 in the \emph{Full range} setup. \emph{(a):} 
The fit with the 3D NLTE model \texttt{A} and \emph{(b):} the respective 1D NLTE 
fit with model \texttt{a}. The four free parameters controlled during the fit are 
highlighted in the plot legend in blue, whereas \vsini\ and the level of 
the continuum were kept fixed. The residuals (dashed lines) are enhanced 
by a factor of ten for better visibility, and offset by $+0.6$ units in flux. 
The uncertainties are the formal 1$\sigma$ internal fitting errors for 
the given S/R\,=\,400 of the CFHT spectrum.}
\label{bestfit}
\end{figure*}

\begin{table*}[htbp]
        \caption{Results with the line list of M12+L14 in the \emph{Full range}
          setup, using 3D model \texttt{A} and 1D~\lhdm\ model \texttt{a}. }
        \label{results}
        \centering
        \begin{tabular}{c|l|r|cccccc}
        \hline\hline\noalign{\smallskip}
        \multicolumn{3}{c|}{Models (\texttt{A}) and (\texttt{a}) - \emph{Full range}} & $\sigma_{\mathrm{T_{eff}}}$ & $\sigma_{\mathrm{\log g}}$ & $\sigma_{\mathrm{[Fe/H
]}}$ & $\sigma_{\mathrm{fit}}$ & $\sigma_{\mathrm{blend}}$ & $\sigma_{\mathrm{cont.}}$   \\
\noalign{\smallskip}\hline\noalign{\smallskip}
        \rule{0pt}{2ex}\multirow{2}{*}{3D-NLTE}     & $A$(Li) & 1.687$\pm$0.114    & 0.111     & 0.018   & 0.010  & 0.002       & 0.007         
& 0.012              \\
                                                    & \iso\ [\%]    & 7.984$\pm$4.431    & 0.124     & 1.228   & 2.811  & 0.708       & 2.338         
& 2.060         \\
\noalign{\smallskip}\hline\noalign{\smallskip}
        \multirow{2}{*}{1D-NLTE}                    & $A$(Li) & 1.670$\pm$0.116    & 0.114     & 0.017   & 0.010  & 0.002       & 0.007         
& 0.007              \\
                                                    & \iso\ [\%]    & 8.863$\pm$4.108    & 0.015     & 1.177   & 3.121  & 0.725       & 1.853         
& 1.338             \\
\noalign{\smallskip}\hline
        \end{tabular}
\tablefoot{The different sources of uncertainty are listed separately to
  identify the quantities that have the strongest effect on the lithium 
  abundance $A$(Li) and the \iso\ ratio (see text for details).}
\end{table*}

\subsection{Best fit}
\label{S5.2}
We cross checked the quality of our fits by visual inspection of the
interpolated synthetic spectra. This, in turn, allows us to better understand
the source of the magnitude of the \chisq.  It is particularly useful when
comparing 3D and 1D results. In \mbox{Fig. \ref{bestfit}} we show the results
obtained with the list of atomic data by M12+L14 (= M12 with the 
\ion{V}{i} line corrected according to \citealt{lawler14}) and model \texttt{A}, 
which best fits the lithium region around \mbox{670.8 nm}, as indicated by 
the \chisq\ in Table\,\ref{VImod} and in the lower panel of 
\mbox{Fig \ref{cfht_res}}.

Panels (a) and (b) of \mbox{Fig. \ref{bestfit}} show the best fit from the 3D
and the 1D~\lhdm\ models, respectively, superimposed on the observed CFHT
spectrum.  We report the final results using the M12+L14 line list 
in \mbox{Table \ref{results}} in the \emph{Full range} setup, because 
including all the blends in the wavelength window is more objective than 
limiting the analysis to the Li part only. 

The errors given in the Table consider six sources of uncertainty: 
$\sigma_{\teff}$ denotes the error due to systematic uncertainty in 
the effective temperature of the input model atmospheres of 
$\Delta\teff=\pm 100$\,K, which is consistent with the \teff\ error 
given by \citet{strassmeier11} ($\pm 70$ K, in their Table 3). 
Having access to the fitting results obtained with the 3D model \texttt{B} and the 1D
\lhdm\ model \texttt{b} of Table \ref{models}, which are 200 K cooler than the master
models \texttt{A} and \texttt{a}, $\sigma_{\teff}$ is defined as the semi-difference of the
corresponding $A$(Li) and \iso\ ratio, respectively. Similarly, 
$\sigma_{\logg}$ and $\sigma_{\rm [Fe/H]}$ are the errors due to a variation of
the surface gravity by $\Delta\logg=\pm0.15$\,dex and of the metallicity by
$\Delta\rm [Fe/H]=\pm0.10$\,dex, respectively. The internal fitting error (shown as error bars in \mbox{Fig.\,\ref{cfht_res}}) is $\sigma_{\rm fit}$,
whereas $\sigma_{\rm blend}$ is the error estimate related to the uncertainty of 
the blend modeling (which is the standard deviation of $A$(Li) and 
\iso, respectively, predicted by line lists G09, M12, and I14). Finally,
$\sigma_{\rm cont}$, the error related to the uncertainty of continuum 
placement, is defined as the semi-difference between the results obtained 
when treating the continuum level as a fixed and a free fitting parameter,
respectively. These statistically independent contributions are summed up 
in quadrature to produce the final error bars given in the third column of 
Table \,\ref{results}.  

Our best estimates of the lithium abundance,
\mbox{$\mathrm{A(Li)_{3D-NLTE}}=1.69\pm0.11$ dex} and
$\mathrm{A(Li)_{1D-NLTE}}=1.67\pm0.12$ dex, are both in excellent agreement
with the earlier determination by \citet{strassmeier11} (\mbox{$1.70\pm0.05$
  dex}) based on equivalent widths. A considerable \lisix\ content is detected
using both 3D and 1D models with values of 8.0\,$\pm$\,4.4\% 
(3D-NLTE) and 8.9\,$\pm$\,4.1\% (1D-NLTE). The similarity of these 
isotopic ratios indicates that the 3D effects are relatively small 
for HD\,123351, as expected from the discussion in Sect.~\ref{synthesis}. 

Some uncertainty remains with
the dominant \ion{Fe}{i}\,+\,\ion{CN}{} feature at \mbox{670.74 nm} that 
causes the enhancement in \chisq\ when including this line in the fitting
range. Curiously, for this line the 1D fit shows a slightly better agreement
with the data (see Fig.\,\ref{bestfit}). We believe that this is not related 
to the quality of the 3D model, but either to a 1D calibrated \loggf\ value 
of this line or to a possible uncertainty in the metallicity of this star. 

We demonstrate in Appendix\,\ref{appendix:B2} that a slight increase 
of the C and/or N abundance leads to an almost perfect fit of the 
\ion{Fe}{i}\,+\,\ion{CN}{} feature without significantly changing the 
derived atmospheric \lisix\ content of HD\,123351.

\section{Discussion}
\label{S6}
As shown by \citet{strassmeier11}, and confirmed in Sect.\,\ref{S22}, our
target star, HD\,123351, has a mass of about 1.2--1.3 solar masses and
is located on the lower part of the red giant branch, probably 
experiencing its first dredge-up. In this phase of evolution, it is expected
to show only a very low lithium abundance at the surface, since the
lithium that survived during the main-sequence evolution is diluted 
by a large factor when the convective envelope deepens 
along the lower RGB. At the same time, the temperature at the base of 
the outer convection zone increases to values above $2.5\times 10^6$\,K,
which leads to further destruction of lithium due to nuclear burning, even 
in standard models without extra mixing. For solar metallicity, the 
first dredge-up reduces the lithium abundance by roughly a factor of 
$100$ with respect to its value at the end of the main sequence 
\citep[e.g.,][Fig.\,5]{sackmann99}.

Since our best model of HD\,123351 has an age of approximately $4.5$ Gyr, its
lithium abundance can be meaningfully compared to that observed in members of
the open cluster M67, whose age is $\approx$\,$4$\,Gyr (e.g.,
\citealt{Demarque_ea:1992}; see also \citealt{Barnes_ea:2016}), provided that
only members in an appropriate evolutionary phase are considered, that is,
subgiants and/or red giant branch stars.

\begin{figure}
\begin{center}
\includegraphics[width=0.49\textwidth]{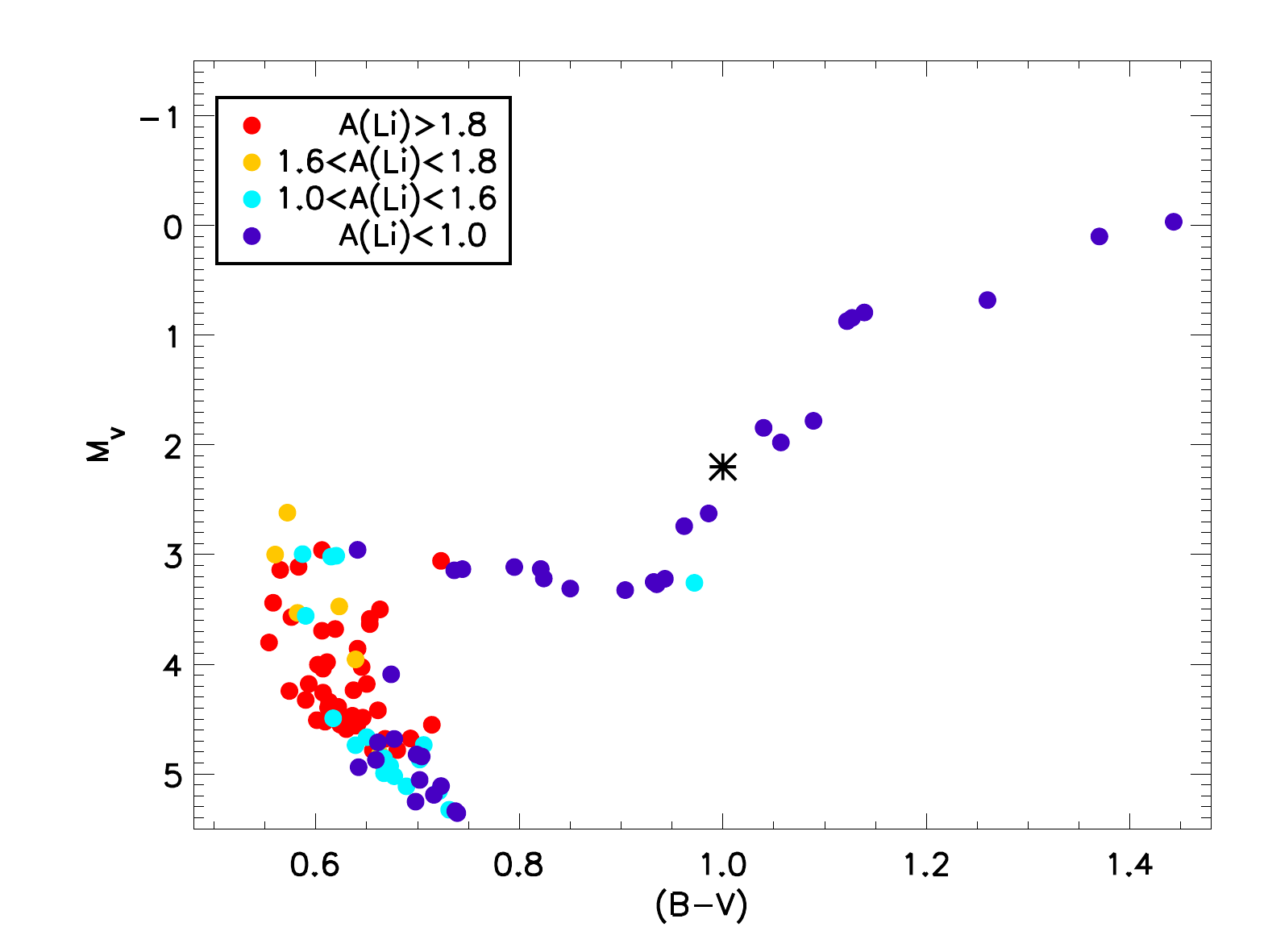}
\caption{CMD of M67 (circles, data from \citealt{Pace_ea:2012}), together 
with HD\,123351 (asterisk). The symbols representing cluster members are 
color-coded to reflect their lithium abundance.
}
\label{m67_comparison}
\end{center}
\end{figure}

In Figure \ref{m67_comparison}, our star is positioned in the color magnitude
diagram (CMD) of M67, where cluster members have been color-coded according
to their Li abundance (based on data from \citealt{Pace_ea:2012}).  It is
clear from this diagram that, in this cluster, a lithium abundance of $A\rm (^7
Li)\lesssim 1.0$ is typical of stars that have evolved past the main sequence
turnoff and populate the red giant branch. This is also in reasonably good
agreement with the results of our standard model discussed in Section
\ref{S22}, where we have argued that the abundance of $A\rm (Li)=1.2$ dex
predicted for our star should be interpreted as an upper limit.

With this background, it is rather surprising that we find a relatively high
lithium abundance of $A$(Li)\,$\approx$\,1.7 in the photosphere of this
star. Apparently, this result calls for a non-standard source of Li
production.  In the following, we discuss possible lithium production
mechanisms which may explain the high content of \liseven\ and \lisix\ found
in HD\,123351.


\subsection{Cameron-Fowler mechanism}

As mentioned in the Introduction, the synthesis of fresh lithium can occur 
in low-mass red giants via the so-called \citet{cameron71} $\rm ^7Be$ 
transport mechanism, if a sufficiently effective deep circulation
connects the convective envelope with the hydrogen-burning shell in the 
stellar interior, as demonstrated by \citet{sackmann99}. In the layers 
that are hot enough to support partial p-p processing, beryllium is 
produced by the nuclear reaction $\rm
^3He(\alpha,\gamma)^7Be$. If mixing is fast enough, part of the $^7$Be is
carried away by the circulation towards cooler regions before the $pp2$
chain can complete. Subsequently, beryllium is converted to \liseven\ by
electron capture in a cooler environment where temperatures are too low for
burning \liseven. Finally, convection efficiently distributes the fresh
lithium throughout the convective envelope, whereby it finally reaches the
surface layers including the stellar photosphere.

In this context, it is worth recalling that HD\,123351 has a binary 
companion in a highly eccentric orbit, which may be responsible for
enhanced tidal interaction during periastron. This, in turn, may provide
the physical mechanism inducing the deep circulation required for the
Cameron-Fowler process to work.

Although this scenario represents a viable way to explain the presence of
a considerable amount of \liseven\ in HD\,123351, the Cameron-Fowler 
mechanism produces only \liseven\ and no \lisix. In view of the rather high
\iso\ isotopic ratio of $\approx 8\%$ derived in this work, it appears
unlikely that the Cameron-Fowler mechanism is the source of Li in our target
star.

\subsection{Stellar flares}

Another source of \lisix\ and \liseven\ production, occurring in the
atmospheric layers of stars, is represented by low energy spallation reactions
with C, N, and O in stellar flares (\citealt{fowler55}, \citealt{canal74}, 
\citealt{canal75}, \citealt{pallavicini92}). The Sun itself seems to be
producing some lithium in flares, as suggested by measurements of the \iso\ 
ratio in the solar wind obtained by analyzing the lunar soil 
($\rm ^6Li/^7Li\approx3\%$, \citealt{chaussidon99}).
According to \citet{ramaty00}, solar flares produce much more \lisix\ than
\liseven\ through the $^3$He induced reaction $^4$He($^3$He,p)$^6$Li.
Assuming that the bulk of the \lisix\ created in the flares at the solar 
surface is transported away by flares or by coronal mass ejections, they find
an order-of-magnitude agreement with the \iso\ measurement in the solar 
wind. 

Even if not all the \lisix\ is expected to be carried away from the 
atmosphere, but is partly retained by the photospheric layers 
of the Sun, this contribution would not be detectable by spectroscopic 
observations, since the additional lithium is quickly diluted within
the mass of the convective envelope: if a large solar flare produces
$10^{30}$ \lisix\ atoms, of which 10\% are deposited in the photosphere,
it would take an enormous number of $10^{14}$ flare events to produce a 
detectable signature in the solar spectrum.  In this sense, the scenario 
proposed by \citet{ramaty00} is also in agreement with the fact that 
no \lisix\ is measured in the photosphere of the Sun. 

The most energetic flaring phenomena are known as superflares and have been
investigated in a sample of 34 solar-type stars by \citet{honda15}. They found
that flares can occur not only in very young pre-MS objects, where the
enhanced rotation and the activity tend to generate superflares, but also 
in older stars with lower \vsini. However, they could not find any empirical
evidence of lithium production by such superflares and no contribution to 
the \lisix\ isotope at all. Other authors have reported a strengthening of 
the lithium line at $\lambda$\,670.8 nm during flares in stars other than the 
Sun, together with an increased \iso\ ratio (\citealt{montes98}, 
\citealt{flores-soriano15}). Clearly, more observations of such phenomena 
are needed to understand whether or not flares are a viable site for 
producing substantial amounts of lithium in stars. 

In the context of Population\,II stars ($M_\ast\approx 0.8$\,M$_\odot$), 
it was estimated by \citet{tatischeff07} that the accumulated effect 
of flare activity during the evolution along the main sequence
(MS; $\approx 10$\,Gyr) can yield significant amounts of \lisix\ (and \liseven).
Under favorable assumptions, the \lisix\ concentration may
reach detectable levels of $^6$Li/H $\approx 10^{-11}$ after several Gyrs
of MS evolution. Although for such reasons the lithium production
via stellar flares is also possible in solar-metallicity stars 
(the nuclear reaction rate is independent of metallicity), it is 
impossible to explain the amount of \lisix\ (and \liseven) detected in
HD\,123351 as being a result of flare production. This is simply because the mass of
the convection zone of this star ($M_{\rm CZ}\approx 1$\,M$_\odot$) is 
larger by orders of magnitude compared that of the metal-poor halo 
stars considered by \citet{tatischeff07}.
Therefore, even though HD\,123351 is a particularly active star, it appears
very unlikely that the lithium detected in its atmosphere is a result of flare
production.
  
\subsection{Accretion of rocky material}

Another possible scenario for explaining the \lisix\ content presented in this
work is linked to the external lithium enrichment being a consequence of the
ingestion of planets, rocky material, or brown dwarfs that have preserved their
initial lithium.  This scenario was first proposed by
\citet{alexander67} considering evolved red giant stars and afterwards
taken up by \citet{israelian01} to explain the \lisix\ content of the
metal-rich solar-type star HD\,82943 \citep{israelian03}. 
\citet{melendez16} studied the solar-twin star HIP\,68468 hosting a 26
$\rm M_{\oplus}$ planet orbiting close to the star, suggesting some planetary
migration from its original position. With an estimated age of 6 Gyr and a
NLTE lithium abundance of $1.52\pm0.03$ dex, this target can be classified as
a good example of lithium enrichment due to planetary ingestion. In this
framework, \citet{melendez16} estimated that a planet with 6 $\rm M_{\oplus}$
is necessary to reproduce the measured lithium abundance, which is higher than
expected for the evolutionary stage of their target star.  

In principle, a similar event could have occurred in HD\,123351 when it expanded
during its ascent along the RGB. The advantage of this scenario is that it
would explain the observed (nearly meteoritic) \iso\ isotopic ratio of about
$8$\%, assuming that the total observed Li content is of planetary origin.
However, we realize that the required amount of accreted planetary matter is 
enormous, since the mass of the convective envelope of HD\,123351 is estimated
to be about one solar mass (containing $\approx$\,$10^{57}$ hydrogen particles)
at its present stage of evolution. To explain the observed 
$A$(Li)\,$\approx$\,1.7, a total of $5\times 10^{46}$ Li particles or 
$5\times 10^{23}$\,g of Li would need to be supplied by the external source. 
Assuming that the fractional lithium abundance of rocky material is 
$1.6\times 10^{-6}$ \citep{McDonough01}, this corresponds to 
$3\times 10^{29}$\,g or $\approx$\,$53$ Earth masses of rocky material. 
An additional constraint is that the material must have been ingested 
recently (during the last few $100$\,Myrs), 
since any previously accreted lithium would have been destroyed in the 
meantime by nuclear reactions at the base of the expanding convective 
envelope.

In view of these tight constraints, the hypothesis of planet engulfment 
or accretion of rocky material appears very unlikely. It is even possible
that the accretion mechanism would not lead to the desired Li enrichment,
but to an additional destruction of lithium \citep[see][]{deal15}.

\subsection{Spurious $^6$Li detection}

Even though we performed the spectroscopic analysis of the Li feature with
great care and attention, we cannot entirely rule out the possibility that
our detection of \lisix\ in HD\,123351 is spurious. For example, a weak blend
that is assigned a too small $\log gf$ value (or is simply missing) in 
all of the line lists would be compensated by an artificial amount of
\lisix. A case in point is the \ion{V}{i} line discussed in 
Sect.\,\ref{sec:VI}.

Moreover, we recall that HD\,123351 is a heavily
spotted star that harbors surface magnetic fields of considerable strength. The
related physics is so far ignored in our model atmospheres and line formation
calculations. Taking these additional complications into account might lead
to subtle changes in the synthetic line profiles and to different values of the
derived \lisix\ abundance. In particular, it is conceivable that the surface
magnetic fields affect the Li feature through Zeeman broadening. In this case,
our non-magnetic models would spuriously overestimate the abundance of \lisix\
to compensate for the missing Zeeman broadening.

\section{Conclusions}\label{S7}

We performed a full 3D NLTE analysis of the lithium resonance line at
\mbox{670.8 nm} in a high-resolution CFHT spectrum of the magnetically active
RGB star HD\,123351. We derived the lithium abundance $A$(Li) and the \iso\ 
isotopic ratio using different fitting techniques, model atmospheres, and a 
sample of blend line lists with atomic and molecular data available in the 
literature. Particularly good fits of the wavelength region bracketing the 
Li doublet were achieved with the line list of \mbox{\citet{melendez12}},
modified to account for the new \loggf\ value measured by 
\citet{lawler14}. It provides the best fit solution
$\mathrm{A(Li)_{3D-NLTE}}$\,=\,1.69\,$\pm$\,0.11\,dex and 
$\rm ^6Li/^7Li$\,=\,8.0\,$\pm$\,4.4\% with a 3D model atmosphere 
whose parameters are in accordance with the earlier analysis by 
\mbox{\citet{strassmeier11}}, \teff/\logg/[Fe/H]\,=\,4780/3.20/0.00. 
The error bars include different sources of systematic uncertainty,
ensuring the robustness of our result. On the other hand, the older line 
list of \mbox{\citet{reddy02}} seems to include erroneous blends that lead 
to non-physical results in terms of a negative \iso\ ratio in HD\,123351. The relatively high \iso\ isotopic ratio agrees with the values found in the
solar-neighborhood interstellar medium (e.g., \citealt{lemoine95}) but
challenges standard stellar evolution theory, which predicts a much lower
\liseven\ abundance and the absence of \lisix\ in an RGB star that experiences
the first dredge-up. 

We have discussed a number of scenarios that might explain the remarkable 
isotopic lithium abundances found in this evolved target. The Cameron-Fowler
mechanism could explain the relatively high \liseven\ abundance of HD\,123351,
where the required extra mixing might be provided by tidal interaction during
periastron passage of the binary companion. However, this scenario cannot
account for the presence of \lisix. A possible source of \lisix\ production
is represented by low energy spallation reactions in stellar flares. Even
though our target star is known for its high level of magnetic activity,
it appears unlikely that a sufficient amount of \lisix\ can be produced by
this mechanism. The planet engulfment hypothesis would naturally explain 
the essentially meteoritic \iso\ isotopic ratio, but would require the 
recent accretion of about 50 Earth masses of rocky material.

This work underlines the need for improved atomic and molecular data to 
establish a complete and reliable identification of the various blends at 
the position of the \lisix\ isotopic absorption line. The present work relies
on published lists of blend lines that have previously been used in Li-related
work, with some modification accounting for recently updated atomic
data for a critical \ion{V}{i} line. 
In contrast to the total Li abundance $A$(Li), the derived \iso\ 
isotopic ratio depends critically on the details of the blending lines.
A weak blend missing in these line lists might easily lead to a
substantial downward revision of the derived \iso\ isotopic ratio.

In a follow-up paper, we plan to analyze very high resolution 
PEPSI\footnote{Potsdam Echelle Polarimetric and Spectroscopic Instrument, mounted at the Large Binocular Telescope (LBT), Arizona (US).} spectra
of HD\,123351 to verify the Li isotopic abundances derived in the present
work from a CFHT spectrum. The PEPSI spectra will also allow us to derive
the \1213C\ isotopic ratio to better constrain the evolutionary state of
HD\,123351. In addition, the PEPSI observations will give us the opportunity 
to study the spectrum of the target at different epochs.

\begin{acknowledgements}
We are grateful to Garik Israelian for kindly sharing his preliminary line list that contributed to the completeness of this work. We would also like to thank 
Piercarlo Bonifacio for providing the SOPHIE spectra of HD\,123351. 
AM thanks the Leibniz-Association for support with a graduate-school grant. 
KGS acknowledges the Canadian CFHT time allocation. 
We thank the anonymous referee for his/her valuable comments and for bringing 
to our attention the new measurement of the \ion{V}{i} atomic data, which 
helped to improve this manuscript. 
This project was supported by FONDATION MERAC.
\end{acknowledgements}

%
%
\bibliographystyle{aa}
\bibliography{references}

\begin{thebibliography}{67}
\expandafter\ifx\csname natexlab\endcsname\relax\def\natexlab#1{#1}\fi

\bibitem[{{Alexander}(1967)}]{alexander67}
{Alexander}, J.~B. 1967, The Observatory, 87, 238

\bibitem[{{Anders} \& {Grevesse}(1989)}]{andersgrevesse89}
{Anders}, E. \& {Grevesse}, N. 1989, \gca, 53, 197

\bibitem[{{Asplund} {et~al.}(2005){Asplund}, {Grevesse}, \&
  {Sauval}}]{asplund05}
{Asplund}, M., {Grevesse}, N., \& {Sauval}, A.~J. 2005, in Astronomical Society
  of the Pacific Conference Series, Vol. 336, Cosmic Abundances as Records of
  Stellar Evolution and Nucleosynthesis, ed. T.~G. {Barnes}, III \& F.~N.
  {Bash}, 25

\bibitem[{{Asplund} {et~al.}(2006){Asplund}, {Lambert}, {Nissen}, {Primas}, \&
  {Smith}}]{asplund06}
{Asplund}, M., {Lambert}, D.~L., {Nissen}, P.~E., {Primas}, F., \& {Smith},
  V.~V. 2006, \apj, 644, 229

\bibitem[{{Barnes} {et~al.}(2016){Barnes}, {Weingrill}, {Fritzewski},
  {Strassmeier}, \& {Platais}}]{Barnes_ea:2016}
{Barnes}, S.~A., {Weingrill}, J., {Fritzewski}, D., {Strassmeier}, K.~G., \&
  {Platais}, I. 2016, \apj, 823, 16

\bibitem[{{Boothroyd} \& {Sackmann}(1999)}]{boothroyd99}
{Boothroyd}, A.~I. \& {Sackmann}, I.-J. 1999, \apj, 510, 232

\bibitem[{{Caffau} \& {Ludwig}(2007)}]{caffauludwig07}
{Caffau}, E. \& {Ludwig}, H.-G. 2007, \aap, 467, L11

\bibitem[{{Cameron} \& {Fowler}(1971)}]{cameron71}
{Cameron}, A.~G.~W. \& {Fowler}, W.~A. 1971, \apj, 164, 111

\bibitem[{{Canal}(1974)}]{canal74}
{Canal}, R. 1974, \apj, 189, 531

\bibitem[{{Canal} {et~al.}(1975){Canal}, {Isern}, \& {Sanahuja}}]{canal75}
{Canal}, R., {Isern}, J., \& {Sanahuja}, B. 1975, \apj, 200, 646

\bibitem[{{Cayrel} {et~al.}(2007){Cayrel}, {Steffen}, {Chand}, {Bonifacio},
  {Spite}, {Spite}, {Petitjean}, {Ludwig}, \& {Caffau}}]{cayrel07}
{Cayrel}, R., {Steffen}, M., {Chand}, H., {et~al.} 2007, \aap, 473, L37

\bibitem[{{Chaboyer} {et~al.}(1995){Chaboyer}, {Demarque}, \&
  {Pinsonneault}}]{chaboyer95}
{Chaboyer}, B., {Demarque}, P., \& {Pinsonneault}, M.~H. 1995, \apj, 441, 876

\bibitem[{{Charbonnel} \& {Balachandran}(2000)}]{charbonnel00}
{Charbonnel}, C. \& {Balachandran}, S.~C. 2000, \aap, 359, 563

\bibitem[{{Charbonnel} \& {Do Nascimento}(1998)}]{charbonnel98}
{Charbonnel}, C. \& {Do Nascimento}, Jr., J.~D. 1998, \aap, 336, 915

\bibitem[{{Chaussidon} \& {Robert}(1999)}]{chaussidon99}
{Chaussidon}, M. \& {Robert}, F. 1999, \nat, 402, 270

\bibitem[{{Deal} {et~al.}(2015){Deal}, {Richard}, \& {Vauclair}}]{deal15}
{Deal}, M., {Richard}, O., \& {Vauclair}, S. 2015, \aap, 584, A105

\bibitem[{{Demarque} {et~al.}(1992){Demarque}, {Green}, \&
  {Guenther}}]{Demarque_ea:1992}
{Demarque}, P., {Green}, E.~M., \& {Guenther}, D.~B. 1992, \aj, 103, 151

\bibitem[{{Demarque} {et~al.}(2008){Demarque}, {Guenther}, {Li}, {Mazumdar}, \&
  {Straka}}]{demarque08}
{Demarque}, P., {Guenther}, D.~B., {Li}, L.~H., {Mazumdar}, A., \& {Straka},
  C.~W. 2008, \apss, 316, 31

\bibitem[{{Denissenkov} \& {Herwig}(2004)}]{Denissenkov2004}
{Denissenkov}, P.~A. \& {Herwig}, F. 2004, \apj, 612, 1081

\bibitem[{{Eggenberger} {et~al.}(2010){Eggenberger}, {Maeder}, \&
  {Meynet}}]{eggenberger10}
{Eggenberger}, P., {Maeder}, A., \& {Meynet}, G. 2010, \aap, 519, L2

\bibitem[{{Flores Soriano} {et~al.}(2015){Flores Soriano}, {Strassmeier}, \&
  {Weber}}]{flores-soriano15}
{Flores Soriano}, M., {Strassmeier}, K.~G., \& {Weber}, M. 2015, \aap, 575, A57

\bibitem[{{Forestini}(1994)}]{forestini94}
{Forestini}, M. 1994, \aap, 285, 473

\bibitem[{{Fowler} {et~al.}(1955){Fowler}, {Burbidge}, \&
  {Burbidge}}]{fowler55}
{Fowler}, W.~A., {Burbidge}, G.~R., \& {Burbidge}, E.~M. 1955, \apjs, 2, 167

\bibitem[{{Freytag} {et~al.}(2012){Freytag}, {Steffen}, {Ludwig},
  {Wedemeyer-B{\"o}hm}, {Schaffenberger}, \& {Steiner}}]{freytag12}
{Freytag}, B., {Steffen}, M., {Ludwig}, H.-G., {et~al.} 2012, Journal of
  Computational Physics, 231, 919

\bibitem[{{Gaia Collaboration} {et~al.}(2016){Gaia Collaboration}, {Brown},
  {Vallenari}, {Prusti}, {de Bruijne}, {Mignard}, {Drimmel}, \&
  {co-authors}}]{gaia}
{Gaia Collaboration}, {Brown}, A.~G.~A., {Vallenari}, A., {et~al.} 2016, ArXiv
  e-prints [\eprint[arXiv]{1609.04172}]

\bibitem[{{Ghezzi} {et~al.}(2009){Ghezzi}, {Cunha}, {Smith}, {Margheim},
  {Schuler}, {de Ara{\'u}jo}, \& {de la Reza}}]{ghezzi09}
{Ghezzi}, L., {Cunha}, K., {Smith}, V.~V., {et~al.} 2009, \apj, 698, 451

\bibitem[{{Grevesse} \& {Sauval}(1998)}]{grevesse98}
{Grevesse}, N. \& {Sauval}, A.~J. 1998, \ssr, 85, 161

\bibitem[{{Honda} {et~al.}(2015){Honda}, {Notsu}, {Maehara}, {Notsu},
  {Shibayama}, {Nogami}, \& {Shibata}}]{honda15}
{Honda}, S., {Notsu}, Y., {Maehara}, H., {et~al.} 2015, \pasj, 67, 85

\bibitem[{{Iben}(1965)}]{iben65}
{Iben}, Jr., I. 1965, \apj, 141, 993

\bibitem[{{Israelian} {et~al.}(2001){Israelian}, {Santos}, {Mayor}, \&
  {Rebolo}}]{israelian01}
{Israelian}, G., {Santos}, N.~C., {Mayor}, M., \& {Rebolo}, R. 2001, \nat, 411,
  163

\bibitem[{{Israelian} {et~al.}(2003){Israelian}, {Santos}, {Mayor}, \&
  {Rebolo}}]{israelian03}
{Israelian}, G., {Santos}, N.~C., {Mayor}, M., \& {Rebolo}, R. 2003, \aap, 405,
  753

\bibitem[{{Klevas} {et~al.}(2016){Klevas}, {Ku{\v c}inskas}, {Steffen},
  {Caffau}, \& {Ludwig}}]{klevas16}
{Klevas}, J., {Ku{\v c}inskas}, A., {Steffen}, M., {Caffau}, E., \& {Ludwig},
  H.-G. 2016, \aap, 586, A156

\bibitem[{{Kupka} {et~al.}(2011){Kupka}, {Dubernet}, \& {VAMDC
  Collaboration}}]{kupka11}
{Kupka}, F., {Dubernet}, M.-L., \& {VAMDC Collaboration}. 2011, Baltic
  Astronomy, 20, 503

\bibitem[{{Kurucz}(1995)}]{kurucz95}
{Kurucz}, R.~L. 1995, \apj, 452, 102

\bibitem[{{Lambert} \& {Ries}(1981)}]{lambert81}
{Lambert}, D.~L. \& {Ries}, L.~M. 1981, \apj, 248, 228

\bibitem[{{Lawler} {et~al.}(2014){Lawler}, {Wood}, {Den Hartog}, {Feigenson},
  {Sneden}, \& {Cowan}}]{lawler14}
{Lawler}, J.~E., {Wood}, M.~P., {Den Hartog}, E.~A., {et~al.} 2014, \apjs, 215,
  20

\bibitem[{{Lemoine} {et~al.}(1995){Lemoine}, {Ferlet}, \&
  {Vidal-Madjar}}]{lemoine95}
{Lemoine}, M., {Ferlet}, R., \& {Vidal-Madjar}, A. 1995, \aap, 298, 879

\bibitem[{{Lind} {et~al.}(2009){Lind}, {Asplund}, \& {Barklem}}]{Lind09}
{Lind}, K., {Asplund}, M., \& {Barklem}, P.~S. 2009, \aap, 503, 541

\bibitem[{{Lind} {et~al.}(2013){Lind}, {Melendez}, {Asplund}, {Collet}, \&
  {Magic}}]{lind13}
{Lind}, K., {Melendez}, J., {Asplund}, M., {Collet}, R., \& {Magic}, Z. 2013,
  \aap, 554, A96

\bibitem[{{Ludwig} {et~al.}(2009){Ludwig}, {Caffau}, {Steffen}, {Freytag},
  {Bonifacio}, \& {Ku{\v c}inskas}}]{ludwig09}
{Ludwig}, H.-G., {Caffau}, E., {Steffen}, M., {et~al.} 2009, \memsai, 80, 711

\bibitem[{{Ludwig} {et~al.}(1994){Ludwig}, {Jordan}, \& {Steffen}}]{ludwig94}
{Ludwig}, H.-G., {Jordan}, S., \& {Steffen}, M. 1994, \aap, 284, 105

\bibitem[{{Magain}(1986)}]{magain86}
{Magain}, P. 1986, \aap, 163, 135

\bibitem[{{Markwardt}(2009)}]{markwardt09}
{Markwardt}, C.~B. 2009, in Astronomical Society of the Pacific Conference
  Series, Vol. 411, Astronomical Data Analysis Software and Systems XVIII, ed.
  D.~A. {Bohlender}, D.~{Durand}, \& P.~{Dowler}, 251

\bibitem[{{McDonough}(2001)}]{McDonough01}
{McDonough}, W.~F. 2001, in International Geophysics Series, Vol.~76,
  Earthquake Thermodynamics and Phase Transitions in the Earth's Interior, ed.
  R.~{Teisseyre} \& E.~{Majewski}, 3--23

\bibitem[{{Melendez} {et~al.}(2016){Melendez}, {Bedell}, {Bean}, {Ramirez},
  {Asplund}, {Dreizler}, {Yan}, {Shi}, {Lind}, {Ferraz-Mello}, {Galarza}, {dos
  Santos}, {Spina}, {Tucci Maia}, {Alves-Brito}, {Monroe}, \&
  {Casagrande}}]{melendez16}
{Melendez}, J., {Bedell}, M., {Bean}, J.~L., {et~al.} 2016, ArXiv e-prints
  [\eprint[arXiv]{1610.09067}]

\bibitem[{{Mel{\'e}ndez} {et~al.}(2012){Mel{\'e}ndez}, {Bergemann}, {Cohen},
  {Endl}, {Karakas}, {Ram{\'{\i}}rez}, {Cochran}, {Yong}, {MacQueen},
  {Kobayashi}, \& {Asplund}}]{melendez12}
{Mel{\'e}ndez}, J., {Bergemann}, M., {Cohen}, J.~G., {et~al.} 2012, \aap, 543,
  A29

\bibitem[{{Montes} \& {Ramsey}(1998)}]{montes98}
{Montes}, D. \& {Ramsey}, L.~W. 1998, \aap, 340, L5

\bibitem[{{M{\"u}ller} {et~al.}(1975){M{\"u}ller}, {Peytremann}, \& {de La
  Reza}}]{mueller75}
{M{\"u}ller}, E.~A., {Peytremann}, E., \& {de La Reza}, R. 1975, \solphys, 41,
  53

\bibitem[{{Nordlund}(1982)}]{nordlund82}
{Nordlund}, A. 1982, \aap, 107, 1

\bibitem[{{Pace} {et~al.}(2012){Pace}, {Castro}, {Mel{\'e}ndez}, {Th{\'e}ado},
  \& {do Nascimento}}]{Pace_ea:2012}
{Pace}, G., {Castro}, M., {Mel{\'e}ndez}, J., {Th{\'e}ado}, S., \& {do
  Nascimento}, Jr., J.-D. 2012, \aap, 541, A150

\bibitem[{{Pallavicini} {et~al.}(1992){Pallavicini}, {Randich}, \&
  {Giampapa}}]{pallavicini92}
{Pallavicini}, R., {Randich}, S., \& {Giampapa}, M.~S. 1992, \aap, 253, 185

\bibitem[{{Pinsonneault}(1997)}]{pinsonneault97}
{Pinsonneault}, M. 1997, \araa, 35, 557

\bibitem[{{Ramaty} {et~al.}(2000){Ramaty}, {Tatischeff}, {Thibaud},
  {Kozlovsky}, \& {Mandzhavidze}}]{ramaty00}
{Ramaty}, R., {Tatischeff}, V., {Thibaud}, J.~P., {Kozlovsky}, B., \&
  {Mandzhavidze}, N. 2000, \apjl, 534, L207

\bibitem[{{Reddy} {et~al.}(2002){Reddy}, {Lambert}, {Laws}, {Gonzalez}, \&
  {Covey}}]{reddy02}
{Reddy}, B.~E., {Lambert}, D.~L., {Laws}, C., {Gonzalez}, G., \& {Covey}, K.
  2002, \mnras, 335, 1005

\bibitem[{{Sackmann} \& {Boothroyd}(1999)}]{sackmann99}
{Sackmann}, I.-J. \& {Boothroyd}, A.~I. 1999, \apj, 510, 217

\bibitem[{{Sbordone} {et~al.}(2010){Sbordone}, {Bonifacio}, {Caffau}, {Ludwig},
  {Behara}, {Gonz{\'a}lez Hern{\'a}ndez}, {Steffen}, {Cayrel}, {Freytag},
  {van't Veer}, {Molaro}, {Plez}, {Sivarani}, {Spite}, {Spite}, {Beers},
  {Christlieb}, {Fran{\c c}ois}, \& {Hill}}]{sbordone10}
{Sbordone}, L., {Bonifacio}, P., {Caffau}, E., {et~al.} 2010, \aap, 522, A26

\bibitem[{{Sbordone} {et~al.}(2014){Sbordone}, {Caffau}, {Bonifacio}, \&
  {Duffau}}]{sbordone14}
{Sbordone}, L., {Caffau}, E., {Bonifacio}, P., \& {Duffau}, S. 2014, \aap, 564,
  A109

\bibitem[{{Schatzman}(1977)}]{schatzman77}
{Schatzman}, E. 1977, \aap, 56, 211

\bibitem[{{Siess} \& {Livio}(1999)}]{siess99}
{Siess}, L. \& {Livio}, M. 1999, \mnras, 308, 1133

\bibitem[{{Somers} \& {Pinsonneault}(2014)}]{somers14}
{Somers}, G. \& {Pinsonneault}, M.~H. 2014, \apj, 790, 72

\bibitem[{{Steffen} {et~al.}(2010){Steffen}, {Cayrel}, {Bonifacio}, {Ludwig},
  \& {Caffau}}]{steffen10}
{Steffen}, M., {Cayrel}, R., {Bonifacio}, P., {Ludwig}, H.-G., \& {Caffau}, E.
  2010, in IAU Symposium, Vol. 265, Chemical Abundances in the Universe:
  Connecting First Stars to Planets, ed. K.~{Cunha}, M.~{Spite}, \&
  B.~{Barbuy}, 23--26

\bibitem[{{Steffen} {et~al.}(2012){Steffen}, {Cayrel}, {Caffau}, {Bonifacio},
  {Ludwig}, \& {Spite}}]{steffen12}
{Steffen}, M., {Cayrel}, R., {Caffau}, E., {et~al.} 2012, Memorie della Societa
  Astronomica Italiana Supplementi, 22, 152

\bibitem[{{Steffen} {et~al.}(2015){Steffen}, {Prakapavi{\v c}ius}, {Caffau},
  {Ludwig}, {Bonifacio}, {Cayrel}, {Ku{\v c}inskas}, \&
  {Livingston}}]{steffen15}
{Steffen}, M., {Prakapavi{\v c}ius}, D., {Caffau}, E., {et~al.} 2015, \aap,
  583, A57

\bibitem[{{Strassmeier} {et~al.}(2000){Strassmeier}, {Washuettl}, {Granzer},
  {Scheck}, \& {Weber}}]{strassmeier00}
{Strassmeier}, K., {Washuettl}, A., {Granzer}, T., {Scheck}, M., \& {Weber}, M.
  2000, \aaps, 142, 275

\bibitem[{{Strassmeier} {et~al.}(2011){Strassmeier}, {Carroll}, {Weber},
  {Granzer}, {Bartus}, {Ol{\'a}h}, \& {Rice}}]{strassmeier11}
{Strassmeier}, K.~G., {Carroll}, T.~A., {Weber}, M., {et~al.} 2011, \aap, 535,
  A98

\bibitem[{{Tatischeff} \& {Thibaud}(2007)}]{tatischeff07}
{Tatischeff}, V. \& {Thibaud}, J.-P. 2007, \aap, 469, 265

\bibitem[{{V{\"o}gler}(2004)}]{vogler04}
{V{\"o}gler}, A. 2004, \aap, 421, 755

\end{thebibliography}


\begin{appendix} 

\section{Atomic and molecular data}\label{appendix:A}  

Tables~\ref{linelist_reddy}, \ref{linelist_ghezzi}, \ref{linelist_melendez},
and \ref{linelist_israelian} present the list of atomic and molecular data
adopted for synthesizing the lithium doublet region. The Li transitions 
accounting for hyper-fine splitting and isotopic shifts are listed in 
Table~\ref{linelist_Li}.

\begin{table}[!htb]
\caption{Line list by \citet{reddy02}.}
\label{linelist_reddy}
\centering
\begin{tabular}{cccc}
\hline\hline
\multicolumn{4}{c}{\textbf{Reddy et al. (2002)}} \\ \hline
Wavelength&Chemical&Excitation&log $gf$\\
$\lambda$ (\AA)&species&potential (eV)&(dex)\\ \hline
6707.381 & CN                    & 1.830 & -2.170 \\ 
6707.433 & \ion{Fe}{i}   & 4.607 & -2.283 \\ 
6707.450 & \ion{Sm}{ii}  & 0.930 & -1.040 \\ 
6707.464 & CN                    & 0.792 & -3.012 \\ 
6707.521 & CN                    & 2.169 & -1.428 \\ 
6707.529 & CN                    & 0.956 & -1.609 \\ 
6707.529 & CN                    & 2.009 & -1.785 \\ 
6707.529 & CN                    & 2.022 & -1.785 \\ 
6707.563 & \ion{V}{i}    & 2.743 & -1.530 \\ 
6707.644 & \ion{Cr}{i}   & 4.207 & -2.140 \\ 
6707.740 & \ion{Ce}{ii}  & 0.500 & -3.810 \\ 
6707.752 & \ion{Ti}{i}   & 4.050 & -2.654 \\ 
6707.771 & \ion{Ca}{i}   & 5.796 & -4.015 \\ 
6707.816 & CN                & 1.206 & -2.317 \\ 
6708.025 & \ion{Ti}{i}   & 1.880 & -2.252 \\ 
6708.094 & \ion{V}{i}    & 1.220 & -3.113 \\ 
6708.125 & \ion{Ti}{i}   & 1.880 & -2.886 \\ 
6708.280 & \ion{V}{i}    & 1.218 & -2.178 \\ 
6708.375 & CN                & 2.100 & -2.252 \\ \hline
\end{tabular}
\end{table}

\begin{table}[!htb]
\caption{Line list by \citet{ghezzi09}.}
\label{linelist_ghezzi}
\centering
\begin{tabular}{cccc}
\hline\hline
\multicolumn{4}{c}{\textbf{Ghezzi et al. (2009)}} \\ \hline
Wavelength&Chemical&Excitation&log $gf$\\
$\lambda$ (\AA)&species&potential (eV)&(dex)\\ \hline
6706.980 & \ion{Si}{i}  & 5.954   & -2.797 \\ 
6707.205 & CN                   & 1.970   & -1.222 \\ 
6707.282 & CN                   & 2.040   & -1.333 \\ 
6707.371 & CN                   & 3.050   & -0.522 \\ 
6707.431 & \ion{Fe}{i}  & 4.608   & -2.268 \\ 
6707.457 & CN                   & 0.790   & -3.055 \\ 
6707.470 & CN                   & 1.880   & -1.451 \\ 
6707.473 & \ion{Sm}{ii} & 0.933   & -1.477 \\ 
6707.518 & \ion{V}{i}   & 2.743   & -1.995 \\ 
6707.545 & CN                   & 0.960   & -1.548 \\ 
6707.595 & CN                   & 1.890   & -1.851 \\ 
6707.596 & \ion{Cr}{i}  & 4.208   & -2.767 \\ 
6707.645 & CN                   & 0.960   & -2.460 \\ 
6707.740 & \ion{Ce}{ii} & 0.500   & -3.810 \\ 
6707.752 & \ion{Sc}{i}  & 4.049   & -2.672 \\ 
6707.771 & \ion{Ca}{i}  & 5.796   & -4.015 \\ 
6707.807 & CN                   & 1.210   & -1.853 \\ 
6707.848 & CN                   & 3.600   & -2.417 \\ 
6707.899 & CN                   & 3.360   & -3.110 \\ 
6707.930 & CN               & 1.980   & -1.651 \\ 
6707.964 & \ion{Ti}{i}  & 1.879   & -6.903 \\ 
6707.980 & CN               & 2.390   & -2.027 \\ 
6708.023 & \ion{Si}{i}  & 6.000   & -2.910 \\ 
6708.026 & CN                   & 1.980   & -2.031 \\ 
6708.094 & \ion{V}{i}   & 1.218   & -3.113 \\ 
6708.147 & CN                   & 1.870   & -1.434 \\ 
6708.275 & \ion{Ca}{i}  & 2.710   & -3.377 \\ 
6708.375 & CN                   & 1.979   & -1.097 \\ 
6708.499 & CN                   & 1.868   & -1.423 \\ 
6708.577 & \ion{Fe}{i}  & 5.446   & -2.728 \\ 
6708.635 & CN                   & 1.870   & -1.584\\ \hline
\end{tabular}
\end{table}

\begin{table}[!htb]
\caption{Line list by \citet{melendez12}.}
\label{linelist_melendez}
\centering
\begin{tabular}{cccc}
\hline\hline
\multicolumn{4}{c}{\textbf{Melendez et al. (2012)}} \\ \hline
Wavelength&Chemical&Excitation&log $gf$\\
$\lambda$ (\AA)&species&potential (eV)&(dex)\\ \hline
6707.000 & \ion{Si}{i}  & 5.954 & -2.560 \\ 
6707.172 & \ion{Fe}{i}  & 5.538 & -2.810 \\ 
6707.205 & CN               & 1.970 & -1.222 \\ 
6707.272 & CN               & 2.177 & -1.416 \\ 
6707.282 & CN               & 2.055 & -1.349 \\ 
6707.300 & C$_{2}$          & 0.933 & -1.717 \\ 
6707.371 & CN               & 3.050 & -0.522 \\ 
6707.433 & \ion{Fe}{i}  & 4.608 & -2.250 \\ 
6707.460 & CN                   & 0.788 & -3.094 \\ 
6707.461 & CN                   & 0.542 & -3.730 \\ 
6707.470 & CN                   & 1.880 & -1.581 \\ 
6707.473 & \ion{Sm}{ii} & 0.933 & -1.910 \\ 
6707.548 & CN                   & 0.946 & -1.588 \\ 
6707.595 & CN                   & 1.890 & -1.451 \\ 
6707.596 & \ion{Cr}{i}  & 4.208 & -2.667 \\ 
6707.645 & CN               & 0.946 & -3.330 \\ 
6707.660 & C$_{2}$              & 0.926 & -1.743 \\ 
6707.809 & CN                   & 1.221 & -1.935 \\ 
6707.848 & CN                   & 3.600 & -2.417 \\ 
6707.899 & CN                   & 3.360 & -3.110 \\ 
6707.930 & CN                   & 1.980 & -1.651 \\ 
6707.970 & C$_{2}$              & 0.920 & -1.771 \\ 
6707.980 & CN                   & 2.372 & -3.527 \\ 
6708.023 & \ion{Si}{i}  & 6.000 & -2.800 \\ 
6708.026 & CN                   & 1.980 & -2.031 \\ 
6708.094 & \ion{V}{i}   & 1.218 & -2.922 \\ 
6708.099 & \ion{Ce}{ii} & 0.701 & -2.120 \\ 
6708.147 & CN                   & 1.870 & -1.884 \\ 
6708.282 & \ion{Fe}{i}  & 4.988 & -2.700 \\ 
6708.315 & CN                   & 2.640 & -1.719 \\ 
6708.347 & \ion{Fe}{i}  & 5.486 & -2.580 \\ 
6708.370 & CN                   & 2.640 & -2.540 \\ 
6708.420 & CN                   & 0.768 & -3.358 \\ 
6708.534 & \ion{Fe}{i}  & 5.558 & -2.936 \\ 
6708.541 & CN                   & 2.500 & -1.876 \\ 
6708.577 & \ion{Fe}{i}  & 5.446 & -2.684 \\ \hline
\end{tabular}
\end{table}

\begin{table}[!htb]
\caption{Line list by Israelian (2014, priv. comm.).}
\label{linelist_israelian}
\centering
\begin{tabular}{cccc}
\hline\hline
\multicolumn{4}{c}{\textbf{Israelian (2014, priv. comm.)}} \\ \hline
Wavelength&Chemical&Excitation&log $gf$\\
$\lambda$ (\AA)&species&potential (eV)&(dex)\\ \hline
6707.010 & \ion{Si}{i}  & 5.954 & -2.784 \\ 
6707.381 & CN                   & 1.830 & -2.170 \\ 
6707.426 & \ion{Fe}{i}  & 4.610 & -2.293 \\ 
6707.450 & \ion{Sm}{ii} & 0.930 & -1.040 \\ 
6707.464 & CN                   & 0.792 & -3.012 \\ 
6707.521 & CN                   & 2.169 & -1.428 \\ 
6707.551 & CN                   & 0.956 & -1.609 \\ 
6707.551 & CN                   & 2.009 & -1.785 \\ 
6707.551 & CN                   & 2.022 & -1.785 \\ 
6707.563 & \ion{V}{i}   & 2.743 & -1.530 \\ 
6707.604 & \ion{Cr}{i}  & 4.208 & -2.410 \\ 
6707.740 & \ion{Ce}{ii} & 0.500 & -3.810 \\ 
6707.752 & \ion{Ti}{i}  & 4.050 & -2.654 \\ 
6707.771 & \ion{Ca}{i}  & 5.800 & -4.015 \\ 
6707.816 & CN                   & 1.206 & -2.317 \\ 
6708.025 & \ion{Si}{i}  & 6.000 & -2.970 \\ 
6708.094 & \ion{V}{i}   & 1.218  & -3.113 \\ 
6708.125 & \ion{Ti}{i}  & 1.880 & -2.886 \\ 
6708.280 & \ion{V}{i}   & 1.220 & -2.178 \\ 
6708.375 & CN                   & 2.100 & -2.252 \\ 
6708.700 & \ion{Fe}{i}  & 4.220 & -3.658 \\ \hline
\end{tabular}
\end{table}

\begin{table}[!htb]
\caption{The twelve components (\liseven\ and \lisix) representing the lithium 
hyper-fine structure (HFS) and isotopic shifts, regrouped from the original sample 
of 19 components given by \citet{kurucz95}.}
\label{linelist_Li}
\centering
\begin{tabular}{cccc}
\hline\hline
\multicolumn{4}{c}{\textbf{Lithium HFS}} \\ \hline
Wavelength&Lithium&Excitation&log $gf$\\
$\lambda$ (\AA)&isotope&potential (eV)&(dex)\\ \hline
6707.756 & \liseven & 0.000 & -0.428 \\ 
6707.768 & \liseven & 0.000 & -0.206 \\ 
6707.907 & \liseven & 0.000 & -0.808 \\ 
6707.908 & \liseven & 0.000 & -1.507 \\ 
6707.919 & \liseven & 0.000 & -0.808 \\ 
6707.920 & \liseven & 0.000 & -0.808 \\ 
6707.920 & \lisix   & 0.000 & -0.479 \\ 
6707.923 & \lisix   & 0.000 & -0.178 \\ 
6708.069 & \lisix   & 0.000 & -0.831 \\ 
6708.070 & \lisix   & 0.000 & -1.734 \\ 
6708.074 & \lisix   & 0.000 & -0.734 \\ 
6708.075 & \lisix   & 0.000 & -0.831 \\ \hline
\end{tabular}
\end{table}

\clearpage 
\section{The chemical composition of HD\,123351}
\subsection{Abundance analysis of SOPHIE spectra}
\label{appendix:B1}

We verified the chemical composition of the star by analyzing three
spectra observed with the SOPHIE spectrograph mounted at the 1.93\,m 
telescope of the Observatory de Haute-Provence. Two spectra were 
observed on 28.01.2009 (S09) and 05.03.2014 (S14) at high efficiency 
(resolving power $R$\,$\sim$\,$40\,000$) and one was taken on 
21.01.2017 (S17) at high spectral resolution ($R$\,$\sim$\,$75\,000$).

\begin{table}[!htb]
\caption{Solar chemical composition used for the spectrum synthesis
         (Col.\,2) and chemical abundances of available elements 
         (Col.\,7) derived with MyGIsFOS from three SOPHIE spectra 
         (S09, S14, S17). Columns (4)--(6) give the number of lines
         analyzed of each spectrum.}
\label{tab:chem}
\centering
 \resizebox{\columnwidth}{!}{
\begin{tabular}{ccccccc}
\hline\hline\noalign{\smallskip}
\multicolumn{7}{c}{Chemical composition HD\,123351}                  \\ \hline\noalign{\smallskip}
Element    & $A \rm (X)_\odot$\tablefootmark{a}  & Ionization &\multicolumn{3}{c}{ N lines}    & $A\rm (X)_\star$        \\ \cline{4-6}    \noalign{\smallskip}
           & [dex]              & stage       & S09 &S14 &S17                 & [dex]         \\ \noalign{\smallskip}\hline \noalign{\smallskip}
C          & 8.39               & I           &  1  &  1   & 1                & $8.46\pm0.15$ \\ 
N          & 7.80               & -           &  -  &  -   & -                & -             \\ 
O          & 8.66               & I           &  -  &  -   & 1                & $8.60\pm0.10$ \\ 
Si         & 7.55               & I           &  5  &  -   & 6                & $7.49\pm0.13$ \\ 
Ca         & 6.36               & I           &  1  &  1   & -                & $6.38\pm 0.15$ \\ 
Sc         & 3.17               & I           &  4  &  4   & 4                & $3.22\pm 0.10$ \\
Ti         & 5.02               & I           & 11  &  9   & 4                & $4.88\pm0.10$ \\
V          & 4.00               & I           &  3  &  3   & 2                & $3.95\pm0.10$ \\
Cr         & 5.67               & I           &  3  &  3   & 3                & $5.62\pm 0.15$ \\
Fe         & 7.50               & I           & 63  & 60   & 44               & $7.44\pm0.10$ \\
Ce         & 1.58               & -           &  -  &  -   & -                & -             \\
Sm         & 1.01               & -           &  -  &  -   & -                & -             \\ 
\hline
\end{tabular}
}
\tablefoot{\tablefoottext{a}{\citet{grevesse98}, except for CNO taken from 
\citet{asplund05}.}}
\end{table}

Individual chemical abundances were derived with the automated analysis 
pipeline MyGIsFOS \citep{sbordone14}, assuming the stellar parameters 
\mbox{$\teff=4780$ K}, \mbox{$\logg=3.2$}, and 
\mbox{$\xi_{\rm micro} =1.4$\,\kms}. We report the resulting abundances 
of importance for the blends in the \emph{Li range} in Table\,\ref{tab:chem}.
The carbon abundance of \mbox{$A({\rm C})=8.46\pm 0.15$} is based on the 
analysis of a single atomic line at 538.0\,nm.
We also investigated the G-band, but could not determine a reliable carbon
abundance from this molecular band because the CH lines are saturated and 
the region is contaminated by saturated atomic lines, which makes the
definition of the continuum difficult.
The oxygen abundance of \mbox{$A({\rm O})=8.60\pm 0.10$} is derived from 
the forbidden [OI] line.  
Unfortunately, we could not derive a nitrogen abundance since the relevant
atomic lines are located in the infrared, outside the SOPHIE spectral
range.

Nevertheless, for all the major elements whose lines are present in the
various blends around the lithium doublet, our chemical analysis indicates
that, within the error bars, the metallicity of this star is consistent with a
solar chemical composition, thus confirming the value of $\rm [Fe/H]=0.00\pm0.08$
obtained by \citet{strassmeier11}. We also confirm that their stellar
parameters, adopted in this work, yield an optimal consistency of excitation 
and ionization equilibria.

Throughout this work, we used the CIFIST2006 solar abundances
(Col.\,2 of Table\,\ref{tab:chem}) as the input chemical composition 
for the spectrum synthesis calculations. The above analysis shows that 
this choice is fully justified.

\subsection{Fine-tuning the fit of the \ion{Fe}{i}\,+\,\ion{CN}{} blend}
\label{appendix:B2}

As discussed in Section \ref{S5.2}, the quality of the fit of the dominant 
\ion{Fe}{i}\,+\,\ion{CN}{} blend around $\lambda\,670.74$\,nm is relatively 
poor. Here we describe some further tests to investigate to what extent 
an improved fit of this feature would affect our \lisix\ detection in 
HD\,123351.

From the best fits shown in Fig.\,\ref{bestfit}, it is obvious that some
opacity is missing in the vicinity of the \ion{Fe}{i}+\ion{CN}{} blend
at $\lambda\,670.74$\,nm. Considering the line list M12 with the updated 
\ion{V}{i} line at $\lambda\,670.81$\,nm, we additionally experimented
with changing the \loggf\ value of the dominant \ion{Fe}{i} line at 
$\lambda\,670.7433$\,nm. We recomputed the grid of 1D NLTE synthetic 
spectra, increasing \loggf\ of this particular line by different amounts.
Following this procedure, a reasonable fit of the \ion{Fe}{i} feature 
was achieved with $\Delta\loggf$\,$=$\,$+0.07$ dex. At the same time, 
the best fit required a significant redshift of the whole synthetic
spectrum. This effect was expected, as previously discussed in Section
\ref{S5}, and leads to a reduction of the \iso\ isotopic ratio by about 2.6
percentage points with respect to the values listed in Table
\ref{VImod}. However, we also noticed that increasing the strength of the
\ion{Fe}{i} line could not fully compensate for some missing opacity near the
position of the CN blend. Even though the overall fit of the \emph{Full
  range} is significantly improved by enhancing the \ion{Fe}{i} absorption, 
the fit to the lithium line profile (\emph{Li range}) deteriorates. 
Therefore, we looked for an alternative solution.

\begin{figure}[]
\centering
\resizebox{\hsize}{!}{\includegraphics[clip=true]{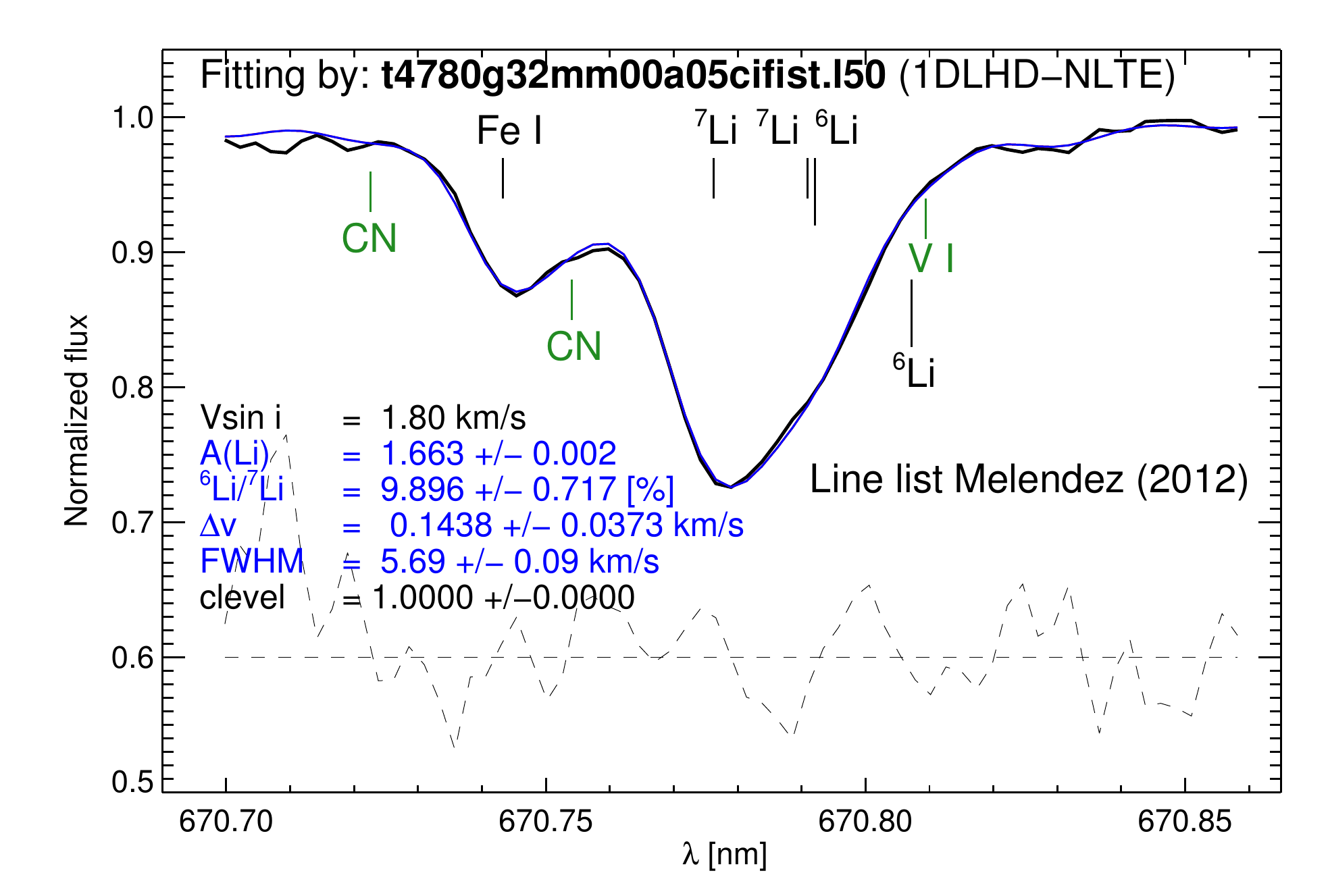}}
\caption{
1D NLTE best-fit of the CFHT spectrum using the 1D \lhdm\ model \texttt{a} of 
Table\,\ref{models} and the list of atomic and molecular data M12, modified to
account for the new \loggf\ value measured by \citet{lawler14} for the 
\ion{V}{i} blend line. For the synthetic spectra, we assumed an enhancement 
in $A$(C) by $+0.06$\,dex with respect to the solar value. The wavelength
positions of the \ion{V}{i} line and of the main contribution to the 
\ion{CN}{} band are indicated in green.}
\label{Cmod}
\end{figure}

As clearly shown in Fig. \ref{linelists}, the CN blend at 670.755 nm should not
to be considered less important than the dominant iron line, since it
contributes significantly to the opacity on the blue part of the \ion{Li}{i}
doublet. Manipulating the \loggf\ values of the CN band is not as convenient 
as for the \ion{Fe}{i} case, since there are several CN lines in this region. 
We instead applied some small changes in carbon abundance $A \rm (C)$.
Considering the line list M12 with the updated \ion{V}{i} line
and recomputing once again the grid of 1D NLTE synthetic spectra with 
different $A$(C), we found a best fit assuming $A \rm (C) = 8.45$ 
(Fig. \ref{Cmod}). This carbon abundance is $0.06$\,dex larger than
the solar value ($A \rm (C)$\,$=$\,$8.39$) adopted in all our previous
spectrum synthesis calculations, and perfectly in agreement with the carbon
abundance derived from the atomic line at $\lambda\,538.0$\,nm
(Appendix\,\ref{appendix:B1}). Alternatively, the same result is obtained when
increasing the abundance of both carbon and nitrogen by $0.04$\,dex.

With this fine-tuning of the carbon abundance, we were not only able to improve
the fit of the CN feature ($\lambda$670.755 nm, see Fig.\,\ref{Cmod}) but
also to match the nearby \ion{Fe}{i} feature almost perfectly. 
The improvement of the fit is a consequence of the fact that by adjusting the
CN blend, the required broadening, expressed by the free parameter $FWHM$, is
lower, leading to a narrower iron feature which naturally reproduces the
observed data. With this slight change of the chemical setup, the measured 
\iso\ ratio is hardly affected (increasing from $8.9$ to $9.9$\%), 
corroborating our previous results.

This analysis could in principle be repeated in 3D NLTE, but
with a much higher computational cost. From Fig. \ref{bestfit} (panel $a$) it
appears that in 3D the opacity missing is larger than in 1D. Presumably,
a 3D fit of the same quality as achieved in 1D would require a slight adjustment
of both the \loggf\ value of the \ion{Fe}{i} line and an increase of the
C and/or N abundance, with all changes well within the uncertainties.
Extrapolating from the 1D experience, it seems reasonable to say that 
the \iso\ ratio is only marginally affected by 
the fine-tuning of the \ion{Fe}{i}\,+\,\ion{CN}{}
blend, thus leaving our 3D NLTE \lisix\ detection of $\approx 8$\% 
unchallenged.

\end{appendix}
\end{document}